# DIFFEOLOGY AND ARITHMETIC OF IRRATIONAL TORI

PATRICK IGLESIAS-ZEMMOUR

ABSTRACT. The irrational torus, $T_\alpha$, originally introduced as a geometric model for quasicrystals, is a foundational object in the theory of diffeology. This paper, after recalling its main algebraic properties, provides a comprehensive analysis of a new geometric invariant for this singular space: the group of flows, $\mathbf{Fl}(T_\alpha, \mathbf{R})$. This invariant, which is trivial for all manifolds, arises as the core of the obstruction to the de Rham theorem in the diffeological setting. We provide a complete computation and geometric interpretation of this group, proving the isomorphism $\mathbf{Fl}(T_\alpha, \mathbf{R}) \simeq \mathbf{R} \times \operatorname{coker}(\Delta_\alpha)$. This result establishes a direct link between the intrinsic geometry of the irrational torus and the deep arithmetic (Diophantine) properties of its defining slope.

## INTRODUCTION

Many of the most fruitful disciplines in mathematics have their roots in physics problems, a tradition that stretches back to Euclidean geometry. The work in this paper follows that same path. The discovery of quasicrystals in the early 1980s presented a profound challenge to classical geometry, demanding new frameworks to describe the aperiodic order they exhibited [SBGC84]. Our work on diffeology, beginning with the 1983 study of what we named the "irrational torus," was a direct response to this challenge [DIZ83]. The goal was to develop a geometry capable of handling the singular quotient spaces that arise as the "internal spaces" in the cut-and-project models of these new physical structures.

Once introduced, however, the irrational torus took on a life of its own as a mathematical object. As is often the case in mathematical physics, the initial physical motivation opened up a new domain of pure geometric inquiry. The irrational torus, with its trivial topology but rich, non-trivial smooth structure, became a foundational example in the developing theory of diffeology —a perfect test case for exploring the properties of singular spaces.

This paper is a contribution to that mathematical exploration. Recent developments in the theory, particularly the construction of the Čech-de Rham bicomplex for diffeological spaces [PIZ24], have provided the necessary tools to revisit this foundational

*Date*: Typeset Jan. 2, 2026.

2020 *Mathematics Subject Classification.* 58A40, 55R15, 37E10, 11J71, 57R30.

I thank the Hebrew University of Jerusalem, Israel, for its continuous academic support. I am also grateful to Professors Hillel Furstenberg and Benjamin Weiss for spending time with me in the past to explain the dynamics of Diophantine approximations.

object and analyze its deeper properties. This has led to the definition of a new geometric invariant, the group of flows $\mathbf{Fl}(\mathrm{X},\mathbf{R})$. This invariant is trivial for manifolds but is designed to be sensitive to the fine smooth structure of singular spaces like $\mathrm{T}_\alpha$; it is the very object that lies at the core of the de Rham obstruction in this new geometry.

This paper provides the complete theoretical foundation and detailed proofs for the results announced in [PIZ25a]. We will:

(1) Formally define the abelian group structure of $\mathbf{Fl}(\mathrm{X},\mathbf{R})$ for any diffeological space X.
(2) Prove the main classification theorem for the irrational torus: $\mathbf{Fl}(\mathrm{T}_\alpha,\mathbf{R}) \simeq \mathbf{R} \times \mathrm{coker}(\Delta_\alpha)$. This includes a detailed analysis of the associated cohomological equation.
(3) Provide a complete geometric interpretation of the components of this group, the "drift" and the "fluctuation," showing how they determine whether the total space of the corresponding bundle is a manifold or an "exotic" non-manifold diffeological space.

The result brings the story full circle, but on a mathematical plane. It demonstrates that the intrinsic geometry of the irrational torus retains a precise memory of the deep arithmetic of its defining slope, confirming its status as a remarkably rich and subtle mathematical object.

## I. Irrational Tori and Algebraic Number Fields

Before introducing the new arithmetic invariant that is the main subject of this paper, we recall the algebraic structures that were discovered in the initial study of the irrational torus. These early results were the first indication that the geometry of this singular space—a model for the quasiperiodic properties of a quasicrystal, just as a regular torus models the periodicity of a crystal—was far richer than its trivial topology would suggest. They revealed that the diffeology of the irrational torus retains a precise memory of the algebraic properties of its defining slope.

For instance, the computation of the group of components of the group of diffeomorphisms of $\mathrm{T}_\alpha = \mathrm{T}^2/\mathscr{S}_\alpha$,[1] denoted $\mathrm{Diff}(\mathrm{T}_\alpha)$, reveals that:

$$\pi_0(\mathrm{Diff}(\mathrm{T}_\alpha)) = \begin{cases} \{\pm 1\} \times \mathbf{Z} & \text{if } \alpha \text{ is a quadratic number} \\ \{\pm 1\} & \text{otherwise.} \end{cases}$$

We recall that a quadratic number is a solution of a quadratic equation with integer coefficients. This relation has been extended to the general case of a codimension 1 linear foliation of a torus, that is, the quotient $\mathrm{T}_\mathrm{H}$ of a torus $\mathrm{T}^n$ by an *irrational hyperplane* H [IZL90]. Precisely, let $\mathrm{T}^n$ be the quotient of $\mathbf{R}^n$ by the lattice $\mathbf{Z}^n$, and $\mathrm{H} \subset \mathbf{R}^n$ be the hyperplane

$$\mathrm{H} = \ker \boldsymbol{\alpha}, \quad \text{with} \quad \boldsymbol{\alpha} = (\alpha_1 = 1, \alpha_2, \ldots, \alpha_n),$$

[1] $\mathscr{S}_\alpha = \{(x, \alpha x) \mod 1 \mid x \in \mathbf{R}\}$ is the irrational 1-parameter subgroup of irrational slope $\alpha$. I chose the notation $\mathscr{S}_\alpha$ to evoke a *solenoid*, by analogy with a toroidal solenoid in electrical engineering, where a wire wound around a toroid traces a similar dense, uniform path.

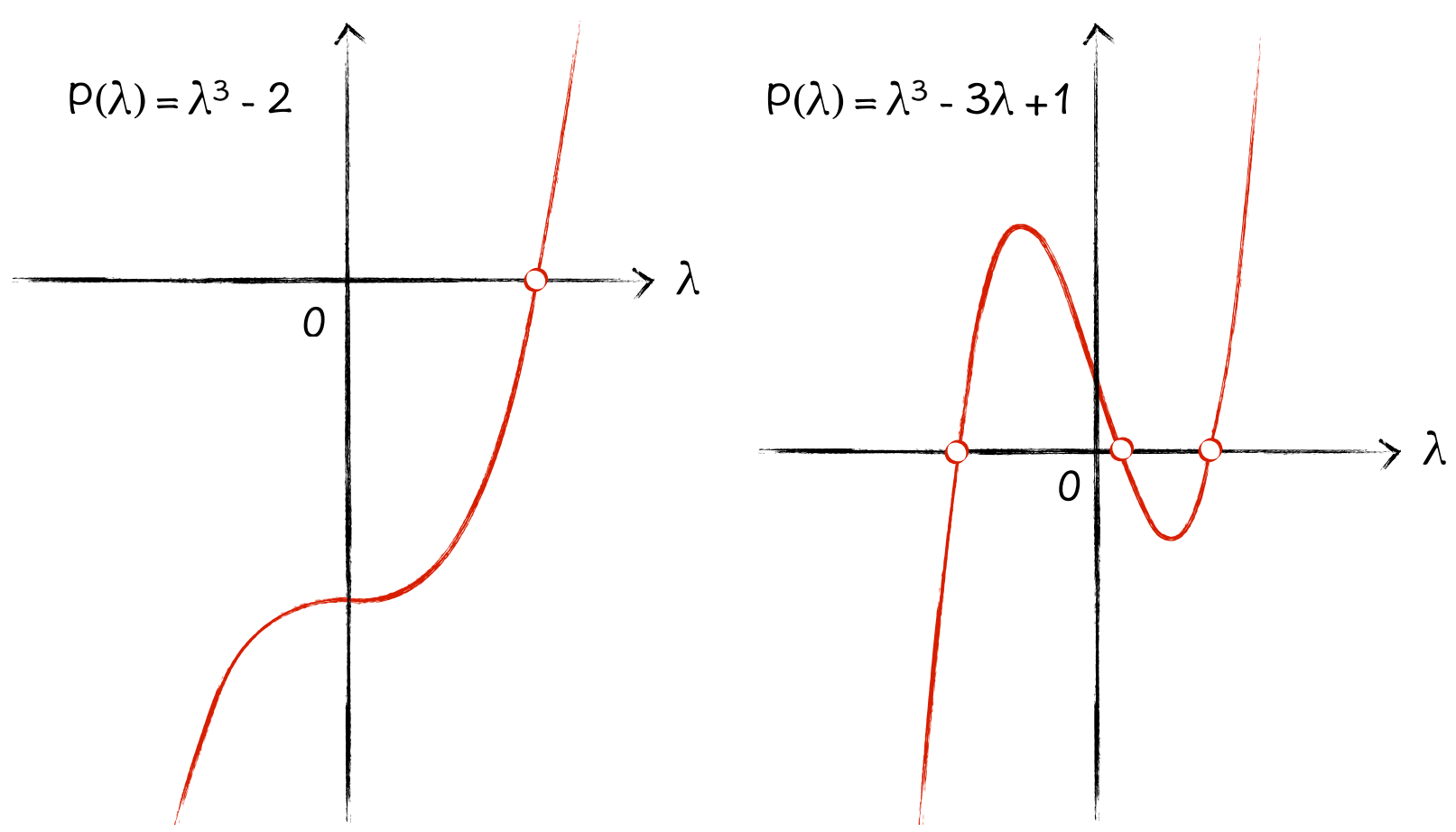

FIGURE 1. The two cases of characteristic field $\mathbf{K}_\alpha$ when $n = 3$.

where $\alpha$ is a linear 1-form on $\mathbf{R}^n$, and the $\alpha_i$ are real numbers independent over $\mathbf{Q}$:

$$\sum_{i=1}^{n} q_i \alpha_i = 0, \ \text{with } q_i \in \mathbf{Q} \quad \Rightarrow \quad q_i = 0.$$

We also denote by $\mathrm{H} \subset \mathrm{T}^n$ the image of H by the standard projection $\mathrm{pr} : \mathbf{R}^n \to \mathrm{T}^n$, i.e., $\mathrm{pr}(x) = \left(e^{2\pi i x_1}, \ldots, e^{2\pi i x_n}\right)$, and let

$$\mathrm{T}_\alpha = \mathrm{T}^n/\mathrm{H}.$$

The irrationality of the hyperplane H (so-called because its director vector $\alpha$ has rationally independent components: $\mathrm{H} \cap \mathbf{Z}^n = \{0\}$.) ensures that its image is dense in $\mathrm{T}^n$, leading to a non-Hausdorff quotient space $\mathrm{T}_\alpha$ with a trivial D-topology, i.e., its only open subsets are $\varnothing$ and $\mathrm{T}_\alpha$. But its diffeology is not trivial and one of its nontrivial aspect is revealed by the computation of the group of components of its group of diffeomorphisms. Here are the objects involved in its computation,

$$\mathrm{E}_\alpha = \alpha(\mathbf{Q}^n) = \left\{ \sum_{i=1}^{n} q_i \alpha_i \mid q_i \in \mathbf{Q} \right\} \quad \text{and} \quad \mathbf{K}_\alpha = \{\lambda \in \mathbf{R} \mid \lambda \mathrm{E} \subset \mathrm{E}\}.$$

The set $\mathrm{E}_\alpha$ is a $\mathbf{Q}$-vector subspace of dimension $n$ in $\mathbf{R}$, it captures the "frequencies" of the irrational foliation. The set $\mathbf{K}_\alpha$ is the set of stabilizers of $\mathrm{E}_\alpha$, by multiplication, representing the scaling symmetries of the frequencies $\mathrm{E}_\alpha$.

**Theorem** (1-IZL). *The set* $\mathbf{K}_\alpha$ *is an algebraic extension of* $\mathbf{Q}$ *and* $\mathrm{E}_\alpha$ *is a* $\mathbf{K}_\alpha$*-vector space. In particular, the dimension* $d$ *of* $\mathbf{K}_\alpha$ *divides* $n$*. The extension* $\mathbf{K}_\alpha$ *is called the* characteristic field *of the space* $\mathrm{T}_\alpha$.

That is, $\mathbf{K}_\alpha = \mathbf{Q}(\lambda)$ for some algebraic number $\lambda$ of degree $d$. And we get the following theorem that highlights how diffeology captures arithmetic invariants within the smooth structure of singular spaces arising from irrational foliations, Establishing a fundamental algebraic structure underlying the geometry of irrational tori.

**Theorem** (2-IZL). *The group* $\pi_0(\mathrm{Diff}(\mathrm{T}_\alpha))$*, of components of the group of diffeomorphisms of* $\mathrm{T}_\alpha$ *is isomorphic to the group of units of an order in the characteristic field* $\mathbf{K}_\alpha$*. Thanks* *to Dirichlet's theorem,*[2] *it is isomorphic to*

$$\pi_0(\mathrm{Diff}(\mathrm{T}_{\boldsymbol{\alpha}})) \simeq \{\pm 1\} \times \mathbf{Z}^{r+s-1},$$

*where $r$ is the number of real roots of the characteristic polynomial* $\mathrm{P}(\lambda)$ *of the field* $\mathbf{K}_{\boldsymbol{\alpha}}$ *and* $2s$ *the number of complex non real solutions.*

As an example the case of $\boldsymbol{\alpha} = (1, \alpha)$, the ordinary irrational torus: $n = 2$, $d$ divides $n$, thus $d = 1$ or $d = 2$. If $d = 1$ then $\mathbf{K}_\alpha = \mathbf{Q}$, $r = 1$, $s = 0$ and $\pi_0(\mathrm{Diff}(\mathrm{T}_\alpha)) \simeq \{\pm 1\}$, otherwise $d = 2$ then $\mathbf{K}_{\boldsymbol{\alpha}} = \mathbf{Q}(\sqrt{\alpha})$, $r = 2$, $s = 0$ and $\pi_0(\mathrm{Diff}(\mathrm{T}_\alpha)) \simeq \{\pm 1\} \times \mathbf{Z}$. The factor $\mathbf{Z}$ represents the subgroup of matrices in $\mathrm{GL}(2, \mathbf{Z})$ that stabilize the line[3] in $\mathbf{R}^2$. That is:

$$\begin{aligned}
\begin{pmatrix} 1 & \alpha \end{pmatrix} \begin{pmatrix} c & a \\ d & b \end{pmatrix} &= \begin{pmatrix} c + \alpha d & a + \alpha b \end{pmatrix} \\
&= (c + \alpha d) \begin{pmatrix} 1 & \dfrac{a + \alpha b}{c + \alpha d} \end{pmatrix} \\
&= \lambda \begin{pmatrix} 1 & \alpha \end{pmatrix} \text{ with } \lambda = c + \alpha d.
\end{aligned}$$

This is an Abelian group made of the powers of its generator.

For an hyperplane $\mathrm{H} \subset \mathbf{R}^3$, except the trivial case we have two possible situations, depending on the characteristic polynomial represented in Figure $i$. On left: Field corresponding to the characteristic polynomial $\mathrm{P}(\lambda) = \lambda^3 - 2$ (one real root, $r = 1$, $s = 1$), leading to $\pi_0(\mathrm{Diff}(\mathrm{T}_{\boldsymbol{\alpha}})) \simeq \{\pm 1\} \times \mathbf{Z}^{1+1-1} = \{\pm 1\} \times \mathbf{Z}$. On right: Field corresponding to $\mathrm{P}(\lambda) = \lambda^3 - 3\lambda + 1$ (three real roots, $r = 3$, $s = 0$), leading to $\pi_0(\mathrm{Diff}(\mathrm{T}_{\boldsymbol{\alpha}})) \simeq \{\pm 1\} \times \mathbf{Z}^{3+0-1} = \{\pm 1\} \times \mathbf{Z}^2$. Here again, It's an Abelian group made of the products of its generators. There can be more than one depending on the roots of the characteristic polynomial.

## II. $(\mathbf{R}, +)$ Principal Bundles over Diffeological Spaces

In the classical category of differentiable manifold, every fiber bundle with contractible fiber has a global smooth section; see for example [Hus94]. Therefore, every principal bundle over a manifold, with structure group the additive group of real numbers $(\mathbf{R}, +)$, is trivial. However, this does not hold in diffeology. For instance, the irrational torus fibration $\pi : \mathrm{T}^2 \to \mathrm{T}_\alpha$ is a nontrivial $(\mathbf{R}, +)$ principal fibration. See [PIZ85] and [PIZ13, §8.11] for the formal definition of fibration, and principal fibration, in diffeology.

This observation justifies the definition of a new invariant for any diffeological space X: the set $\mathbf{Fl}(\mathrm{X}, \mathbf{R})$ of $(\mathbf{R}, +)$ principal bundles, modulo isomorphisms, with base X. Such fiber bundles are also called *flows* over X, since they are defined by a smooth action of $\mathbf{R}$. By construction, $\mathbf{Fl}(\mathrm{X}, \mathbf{R})$ is a uniquely diffeological invariant, as $\mathbf{Fl}(\mathrm{X}, \mathbf{R})$ contains only the trivial bundle when X is a manifold. It reveals an internal complexity that is not revealed by previously introduced diffeological invariants.

This construction, initially introduced in [PIZ85] and [PIZ86], is summarized below.

### II.1. The Group $\mathbf{Fl}(\mathrm{X}, \mathbf{R})$.

---

[2] For a reference to Dirichlet's theorem see [BC67].

[3] Actually the kernel of the linear 1-form $\boldsymbol{\alpha}$ is the line $y = -x/\alpha$, which is equivalent.

Let $\pi : \mathrm{Y} \to \mathrm{X}$ and $\pi' : \mathrm{Y}' \to \mathrm{X}$ be two $(\mathbf{R},+)$-principal bundle. Let us define the following operation, described in two steps:

1. Let $\pi \otimes \pi' : \mathrm{Y} \otimes \mathrm{Y}' \to \mathrm{X}$ be the fiber product defined by
$$\mathrm{Y} \otimes \mathrm{Y}' = \{(y, y') \in \mathrm{Y} \times \mathrm{Y}' \mid \pi(y) = \pi'(y')\},$$
with $\pi \otimes \pi'(y, y') = \pi(y) = \pi'(y')$. Note that the total space $\mathrm{Y} \otimes \mathrm{Y}'$ is identical to the pullback $\pi^*(\mathrm{Y}')$. The result bundle $\pi \otimes \pi'$ is an $(\mathbf{R}^2,+)$ principal bundle for the product of the action on each factor, i.e.,
$$(t, t')_{\mathrm{Y}\otimes\mathrm{Y}'}(y, y') = (t_{\mathrm{Y}}(y), t'_{\mathrm{Y}'}(y')).$$
Let $\overline{\mathbf{R}}$ be the antidiagonal:
$$\overline{\mathbf{R}} = \{(t, -t) \mid t \in \mathbf{R}\} \subset \mathbf{R}^2,$$
It is a subgroup of $(\mathbf{R}^2,+)$.
2. Let $\mathrm{Y}''$ be the quotient $\mathrm{Y}'' = (\mathrm{Y} \otimes \mathrm{Y}')/\overline{\mathbf{R}}$ with $\pi'' : \mathrm{Y}'' \to \mathrm{X}$, such that, for all $y'' = [y, y']$, $\pi''(y'') = \pi(y) = \pi'(y')$, where $[y, y']$ denotes the orbit of the pair $(y, y')$ by the action of $\overline{\mathbf{R}}$. That is:
$$[y, y'] = \{(t_{\mathrm{Y}}(y), -t_{\mathrm{Y}'}(y')) \mid t \in \mathbf{R}\}.$$

This construction leads to the following propositions, on the properties of this operation.

**Proposition** (A). *The projection $\pi'' : \mathrm{Y}'' \to \mathrm{X}$ is again a $(\mathbf{R},+)$-principal bundle for the additive action of $\mathbf{R}$, defined by:*
$$s_{\mathrm{Y}''}[y, y'] = [s_{\mathrm{Y}}(y), y'] = [y, s_{\mathrm{Y}'}(y')].$$

**Proposition** (B). *The operation $(\pi, \pi') \mapsto \pi''$ passes to the equivalence classes of $(\mathbf{R},+)$-principal bundle over $\mathrm{X}$. This operation is denoted additively:*
$$\mathrm{class}(\pi) + \mathrm{class}(\pi') = \mathrm{class}(\pi'').$$
*Let $p = \mathrm{class}(\pi)$, $p' = \mathrm{class}(\pi')$, and $p'' = \mathrm{class}(\pi'')$: the operation $(p, p') \mapsto p'' = p + p'$ is an Abelian group operation on $\mathbf{Fl}(\mathrm{X}, \mathbf{R})$:*

(a) *This operation is Abelian $p + p' = p' + p$.*
(b) *This operation is associative $(p + p') + p'' = p + (p' + p'')$.*
(c) *The class of the trivial bundle $\mathrm{pr}_1 : \mathrm{X} \times \mathbf{R} \to \mathrm{X}$ is the identity element.*
(d) *The inverse $-p$ is the class of the same principal fiber bundle $\pi : \mathrm{Y} \to \mathbf{R}$, but with the inverse $(\mathbf{R},+)$-action $\bar{t}_{\mathrm{Y}}(y) = (-t)_{\mathrm{Y}}(y)$.*

NOTE. The group of flows $\mathbf{Fl}(\mathrm{X}, \mathbf{R})$ appears also, independently, in the *Čech-de-Rham bi-complex in Diffeology*, as the bi-graded cohomology groups $\mathsf{H}^{1,0}_{\delta}(\mathrm{X})$, see [PIZ24, §21] and here [PIZ88].

**Remark.** This invariant is native to singular geometry. Its vanishing for manifolds highlights the unique contribution of the diffeological framework and reveals a profound hierarchy of geometric structure. A fiber-preserving bijection between the torus $\mathrm{T}^2$ and the product $\mathrm{T}_\alpha \times \mathbf{R}$ can be constructed using the Axiom of Choice. This insight reveals that all the total spaces we are considering—the manifold $\mathrm{T}^2$, the trivial bundle $\mathrm{T}_\alpha \times \mathbf{R}$, and all the exotic spaces—are isomorphic as principal bundles in the category of sets. They are different realizations of the same underlying "principal projection."[4]

The geometric distinctions between them are therefore not about the points or the set-theoretic fibration, but purely about the *smooth structure* (the diffeology) imposed upon them. The group $\mathbf{Fl}(\mathrm{T}_\alpha, \mathbf{R})$ is thus revealed to be a classification of these distinct smooth structures that can be placed on a single, common underlying principal projection. As we will see in detail, the different elements of this group correspond to:

– The zero element, corresponding to the standard product diffeology.
– A one-parameter family, corresponding to the unique *manifold* structures that can be placed on this set, yielding the tori $\mathrm{T}^2$.
– An infinite-dimensional family (for non-Diophantine $\alpha$), corresponding to the "exotic," non-manifold diffeologies.

Diffeology gives concrete geometric embodiment to each of these algebraic classes, providing the language and tools to distinguish between structures that are indistinguishable at the level of sets.

**Proof.** (a) First of all, this operation is Abelian: $\mathrm{Y} \otimes \mathrm{Y}'$ is equivalent to $\mathrm{Y}' \otimes \mathrm{Y}$ and the action of $(t, -t) \in \overline{\mathbf{R}}$ becomes $(-t, t) \in \overline{\mathbf{R}}$, and $[y, y'] \simeq [y', y]$, and $p + p' = p' + p$.
(b) Next, let us prove first that this operation is associative: Consider $[y, [y', y'']]$ on the one hand and $[[y, y'], y'']$ on the other hand:

$$\begin{aligned}
[y,[y',y'']] &= \{(t_{\mathrm{Y}}(y), -t_{\mathrm{Y}'''}[y',y'']) \mid t \in \mathbf{R}\} \quad \text{with} \quad \mathrm{Y}''' = (\mathrm{Y}' \otimes \mathrm{Y}'')/\overline{\mathbf{R}} \\
&= \{(t_{\mathrm{Y}}(y), [-t_{\mathrm{Y}'}(y'), y'']) \mid t \in \mathbf{R}\} \\
&= \{(t_{\mathrm{Y}}(y), \{((s-t)_{\mathrm{Y}'}(y'), -s_{\mathrm{Y}''}(y'')) \mid s \in \mathbf{R}\}) \mid t \in \mathbf{R}\} \\
&\simeq \{(t_{\mathrm{Y}}(y), (s-t)_{\mathrm{Y}'}(y'), -s_{\mathrm{Y}''}(y'')) \mid s, t \in \mathbf{R}\} \\
&\simeq \{(s_{\mathrm{Y}''''}[y,y'], -s_{\mathrm{Y}''}(y'')) \mid s \in \mathbf{R}\} \quad \text{with} \quad \mathrm{Y}'''' = (\mathrm{Y} \otimes \mathrm{Y}')/\overline{\mathbf{R}} \\
&= [[y,y'],y''].
\end{aligned}$$

Thus,

$$\begin{aligned}
p + (p' + p'') &= \mathrm{class}(\{[y,[y',y'']] \mid y \in \mathrm{Y}, y' \in \mathrm{Y}' \text{ and } y'' \in \mathrm{Y}''\}) \\
&= \mathrm{class}(\{[[y,y'],y''] \mid y \in \mathrm{Y}, y' \in \mathrm{Y}' \text{ and } y'' \in \mathrm{Y}''\}) \\
&= (p + p') + p''.
\end{aligned}$$

(c) Next, let us show that the class of the trivial bundle is the identity element. Let us denote $\mathbf{0}_{\mathrm{X}}$ the class of the trivial principal bundle $\mathrm{pr}_1 : \mathrm{X} \times \mathbf{R} \to \mathrm{X}$. For all element $(y, (x, t)) \in \mathrm{Y} \otimes (\mathrm{X} \times \mathbf{R})$, we have $[y, (x, t)] = [-t_{\mathrm{Y}}(y), (x, 0)]$ with $\pi(y) = x$. Thus, $\{[(y, (x, t))] \mid y \in \mathrm{Y}, \pi(y) = x, \text{ and } t \in \mathbf{R}\} \simeq \mathrm{Y}$, and $p + \mathbf{0}_{\mathrm{X}} = p$, where $p = \mathrm{class}(\pi)$.
(d) Finally, let us construct the inverse of a class of a bundle: Let $\bar{\pi} : \bar{\mathrm{Y}} \to \mathrm{X}$ be the same fiber bundle but with the inverse action, denoted by $t_{\bar{\mathrm{Y}}}(y) = -t_{\mathrm{Y}}(y)$. Consider the diagonal map $\Delta : y \mapsto (y, y) \mapsto [y, y]$, from $\mathrm{Y}$ to $(\mathrm{Y} \otimes \bar{\mathrm{Y}})/\overline{\mathbf{R}}$. The pair $(y, y)$ is equivalent to $t_{\mathrm{Y}\otimes\bar{\mathrm{Y}}}(y, y) = (t_{\mathrm{Y}}(y), -t_{\bar{\mathrm{Y}}}(y) = (t_{\mathrm{Y}}(y), t_{\mathrm{Y}}(y))$. Thus, $\Delta(y) = \Delta(t_{\mathrm{Y}}(y))$. Therefore there exists a smooth map $\sigma : \mathrm{X} \to (\mathrm{Y} \otimes \bar{\mathrm{Y}})/\overline{\mathbf{R}}$ such that $\Delta = \sigma \circ \pi$, and $\pi' \circ \sigma = \mathbf{1}_{\mathrm{X}}$.

[4] A free action of a group on a set makes the set a product of its quotient by the group; the real question is: is it compatible with the structure (here the diffeology)?

$$
\begin{array}{ccc}
\mathrm{Y} & \xrightarrow{\ \Delta\ } & (\mathrm{Y} \otimes \bar{\mathrm{Y}})/\overline{\mathbf{R}} \\
\pi \big\downarrow & \overset{\sigma}{\nearrow} & \big\downarrow \pi' \\
\mathrm{X} & \xrightarrow[\ \mathbf{1}_{\mathrm{X}}\ ]{} & \mathrm{X}
\end{array}
$$

The map $\sigma$ is a smooth section of the $(\mathbf{R}, +)$ principal fiber bundle $\pi' : (\mathrm{Y} \otimes \bar{\mathrm{Y}})/\overline{\mathbf{R}} \to \mathrm{X}$, hence it is trivial [PIZ13, §8.12], and $\bar{p} = \mathrm{class}(\bar{\pi})$ is the inverse of $p$: $p + \bar{p} = \mathbf{0}_{\mathrm{X}}$.

This concludes the proof that $\mathbf{Fl}(\mathrm{X}, \mathbf{R})$, equipped with this operation, is an Abelian group. ►

## II.2. The Group $\mathbf{Fl}(\mathrm{T}, \mathbf{R})$

Let us illustrate this construction by detailing the group $\mathbf{Fl}(\mathrm{T}, \mathbf{R})$ for $\mathrm{T} = \mathbf{R}/\mathrm{K}$, where $\mathrm{K} \subset \mathbf{R}$ is a strict subgroup. We refer to T as a *general* 1*-dimensional torus*, noting that it is irrational when K has more than one generator.

### (A) The cohomology of K with values in $\mathscr{C}^\infty(\mathbf{R})$

The purpose of this section is to identify the group $\mathbf{Fl}(\mathrm{T}, \mathbf{R})$, of equivalence classes of $(\mathbf{R}, +)$ principal bundles over T, with a first group of cohomology of K with values in the Abelian group $\mathscr{C}^\infty(\mathbf{R})$ of smooth real functions on $\mathbf{R}$.

**Definition** (The Cocycles $\tau$). *Consider the subgroup* $\mathrm{K} \subset \mathbf{R}$ *acting on* $\mathscr{C}^\infty(\mathbf{R})$ *by translations*

$$(k, f) \mapsto \mathrm{T}_k^*(f) = f \circ \mathrm{T}_k \quad \text{with} \quad (k, f) \in \mathrm{K} \times \mathscr{C}^\infty(\mathbf{R}) \quad \text{and} \quad \mathrm{T}_k(x) = x + k.$$

*According to the standard definition,*[5] *a* 1*-cocycle of the group* K*, with values in* $\mathscr{C}^\infty(\mathbf{R})$*, twisted by the action of* K *on* $\mathscr{C}^\infty(\mathbf{R})$*, is a map*

$$\tau : \mathrm{K} \to \mathscr{C}^\infty(\mathbf{R}) \quad \text{such that} \quad \tau(k + k') = \mathrm{T}_{k'}^*(\tau(k)) + \tau(k'),$$

*for all* $k, k' \in \mathrm{K}$*. Explicitly, for all* $x \in \mathbf{R}$*,*

$$\tau(k + k')(x) = \tau(k)(x + k') + \tau(k')(x).$$

*We can write also* $\tau(k, x)$ *instead of* $\tau(k)(x)$*. We shall denote by*

$$\mathbf{Z}^1(\mathrm{K}, \mathscr{C}^\infty(\mathbf{R})) = \{\tau \in \mathrm{Maps}(\mathrm{K}, \mathscr{C}^\infty(\mathbf{R})) \mid \tau(k + k') = \mathrm{T}_{k'}^*(\tau(k)) + \tau(k'), \quad \text{for all } k, k' \in \mathrm{K}\},$$

*the space of cocycles* $\tau$*, where* $\mathrm{Maps}(\mathrm{A}, \mathrm{B})$ *denotes the set of all maps from* A *to* B.

**Definition** (The Coboundaries $\Delta\sigma$). *A cocycle* $\tau$ *is a coboundary if there exists a function* $\sigma \in \mathscr{C}^\infty(\mathbf{R})$ *such that:*

$$\tau = \Delta\sigma \quad \text{with} \quad \Delta\sigma(k) = \mathrm{T}_k^*(\sigma) - \sigma.$$

*In other words if* $\tau(k)(x) = \sigma(x + k) - \sigma(x)$*. We denote,*

$$\mathbf{B}^1(\mathrm{K}, \mathscr{C}^\infty(\mathbf{R})) = \{\Delta\sigma \mid \sigma \in \mathscr{C}^\infty(\mathbf{R})\}.$$

[5] For example in [Kir74].

**Definition** (The First Cohomology Group). *As usual, the first cohomology group of* K *with values in* $\mathscr{C}^\infty(\mathbf{R})$*, twisted by the action on* $\mathscr{C}^\infty(\mathbf{R})$ *by translations, is the quotient:*

$$\mathsf{H}^1(\mathrm{K}, \mathscr{C}^\infty(\mathbf{R})) = \mathsf{Z}^1(\mathrm{K}, \mathscr{C}^\infty(\mathbf{R}))/\mathsf{B}^1(\mathrm{K}, \mathscr{C}^\infty(\mathbf{R})).$$

*Actually, with the vector space structure on* $\mathscr{C}^\infty(\mathbf{R})$*, these groups are also real vector spaces.*

**(B) Lifting on $\mathbf{R}\times\mathbf{R}$ the action of** K **on R.**

The cocycles $\tau \in \mathsf{Z}^1(\mathrm{K}, \mathscr{C}^\infty(\mathbf{R}))$ are used to lift, on the product $\mathbf{R}\times\mathbf{R}$, the action of K on $\mathbf{R}$ by:

$$k : \begin{pmatrix} x \\ t \end{pmatrix} \mapsto \begin{pmatrix} x+k \\ t+\tau(k,x) \end{pmatrix}, \tag{E}$$

for all $(x,t) \in \mathbf{R}\times\mathbf{R}$. Indeed, denote this action by $k_{\mathbf{R}\times\mathbf{R}}$, we have:

$$\begin{aligned} k_{\mathbf{R}\times\mathbf{R}} \circ k'_{\mathbf{R}\times\mathbf{R}}(x,t) &= k_{\mathbf{R}\times\mathbf{R}}\big(k'_{\mathbf{R}\times\mathbf{R}}(x,t)\big) = k_{\mathbf{R}\times\mathbf{R}}(x+k', t+\tau(k',x)) \\ &= (x+k'+k, t+\tau(k',x)+\tau(k,x+k')) \\ &= (x+k+k', t+\tau(k+k',x)) = (k+k')_{\mathbf{R}\times\mathbf{R}}(x,t). \end{aligned}$$

Note also that

$$\tau(0,x) = 0 \quad\text{and}\quad \tau(-k,x) = \tau(k,x-k).$$

**(C) $(\mathbf{R},+)$ principal fiber bundle over** T **associated with a cocycle $\tau$.**

Let $\tau \in \mathsf{Z}^1(\mathrm{K}, \mathscr{C}^\infty(\mathbf{R}))$. Denote by

$$\mathbf{R}\times_\tau \mathbf{R} = (\mathbf{R}\times\mathbf{R})/\mathrm{K}$$

the quotient of $\mathbf{R}\times\mathbf{R}$ by the action of K lifted by the cocycle $\tau$, as defined in (E). Let

$$\mathrm{pr} : \mathbf{R}\times\mathbf{R} \to \mathbf{R}\times_\tau\mathbf{R} \quad\text{with}\quad \mathrm{pr}(x,t) = [x,t],$$

the canonical projection. We denote also by $[x]$ the orbit of $x \in \mathbf{R}$ by K, and by

$$\pi : \mathbf{R}\times_\tau\mathbf{R} \to \mathrm{T} = \mathbf{R}/\mathrm{K}, \quad\text{the projection}\quad [x,t] \mapsto [x].$$

**Theorem** (Construction). *Let us denote by* Y *the space* $\mathbf{R}\times_\tau\mathbf{R}$*. For all* $s \in \mathbf{R}$*, let*

$$s_{\mathrm{Y}}[x,t] = [x,t+s].$$

*This defines a smooth free action of* $(\mathbf{R},+)$ *on* Y *that makes* $\pi : \mathrm{Y} \to \mathrm{T}$*, the projection* $[x,t] \to [x]$*, a principal fiber bundle.*

Then, let us denote by H this association:

$$\mathrm{H} : \mathsf{Z}^1(\mathrm{K}, \mathscr{C}^\infty(\mathbf{R})) \to \mathbf{Fl}(\mathrm{T},\mathbf{R}) \quad\text{with}\quad \mathrm{H}(\tau) = \mathrm{class}\big(\pi : \mathrm{Y} = \mathbf{R}\times_\tau\mathbf{R} \to \mathrm{T} = \mathbf{R}/\mathrm{K}\big)$$

We conclude by the main theorem:

**Theorem** (Classification). *The association* $\mathrm{H} : \mathsf{Z}^1(\mathrm{K}, \mathscr{C}^\infty(\mathbf{R})) \to \mathbf{Fl}(\mathrm{T},\mathbf{R})$ *is a homomorphism,*

$$\mathrm{H}(\tau+\tau') = \mathrm{H}(\tau) + \mathrm{H}(\tau').$$

*It is surjective (epimorphism) and its kernel is the subgroup of coboundaries:*

$$\ker(\mathrm{H}) = \mathsf{B}^1(\mathrm{K}, \mathscr{C}^\infty(\mathbf{R})).$$

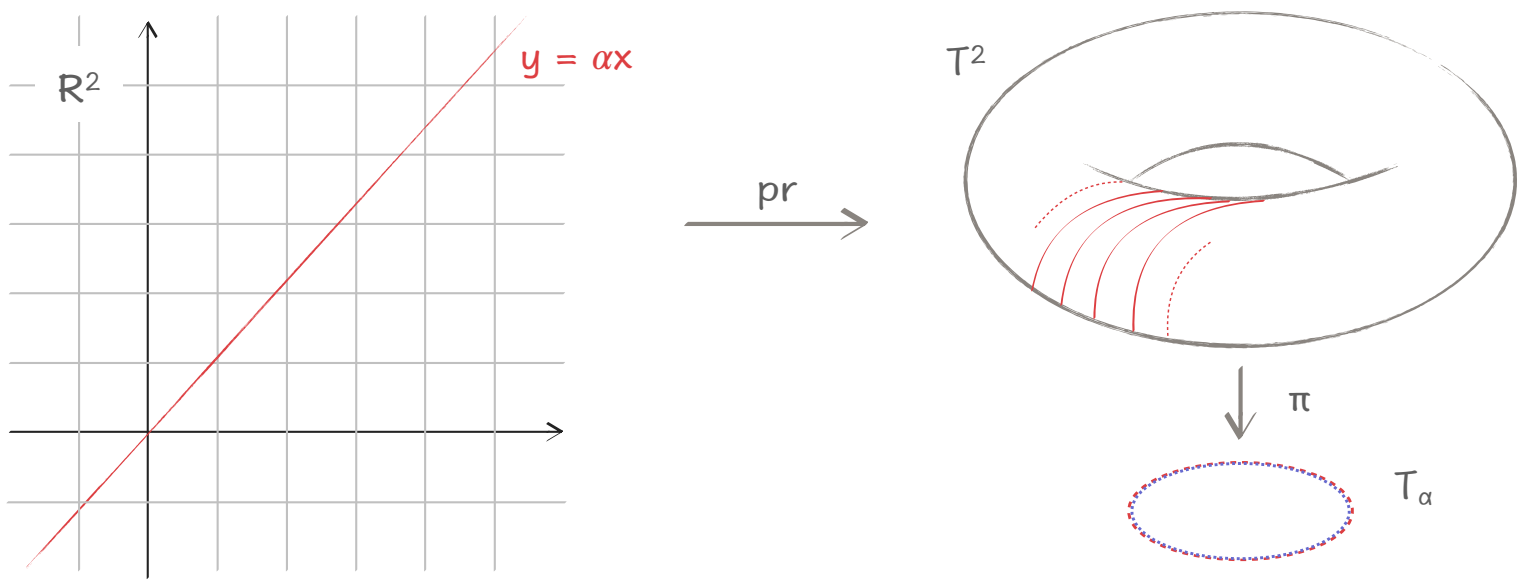

FIGURE II. The Construction of $T_\alpha$.

*Therefore, its projection (still denoted by* H*) is an isomorphism:*

$$\mathbf{Fl}(T,\mathbf{R}) \underset{H}{\simeq} \mathbf{H}^1(K, \mathscr{C}^\infty(\mathbf{R})).$$

**Note 1.** In the construction above, the projection $\mathrm{pr} : (x,t) \mapsto [x,t]$ is the universal covering on $Y = \mathbf{R} \times_\tau \mathbf{R}$, and $\pi_1(Y) = K$. We recover this result by applying the long exact sequence of homotopy to the fiber bundle $\pi : Y \to T$, since the fiber $\mathbf{R}$ is contractible.

**Proof.** We now proceed to prove the theorems stated in the article.

**(1)** First, we verify that the projection $\mathrm{pr} : (x,t) \mapsto [x,t]$ from $\mathbf{R} \times \mathbf{R}$ to $Y = \mathbf{R} \times_\tau \mathbf{R}$ is a covering. Let $P : r \mapsto y_r$ be a plot of Y and consider

$$P^*(\mathbf{R} \times \mathbf{R}) = \{(r,(x,t)) \in \mathrm{dom}(P) \times \mathbf{R} \times \mathbf{R} \mid [x,t] = P(r)\}.$$

Since $(x,t) \mapsto [x,t]$ is a subduction the plot P has a local lift $r \mapsto (x_r, t_r)$. Then, $\Psi'$ : $(r,k) \mapsto (r,(x_r + k, t_r + \tau(k)(x_r)))$ is a smooth bijection from $\mathrm{dom}(P) \times K$ to $P^*(\mathbf{R} \times \mathbf{R})$. Its inverse is given by $\Psi'^{-1}(r,(x,t)) = (r, x - x_r)$ which is clearly smooth. Thus, $\mathrm{pr} : (x,t) \mapsto [x,t]$ is a diffeological fibration with fiber K, thus a covering, and since $\mathbf{R} \times \mathbf{R}$ is simply connected, it is the universal covering [PIZ13, §8.26].

**(2)** Next, let us prove that the projection $\pi : Y = \mathbf{R} \times_\tau \mathbf{R} \to T = \mathbf{R}/K$ is a $(\mathbf{R},+)$ principal fibration, with its action of $\mathbf{R}$ : $s_Y[x,t] = [x, t+s]$. Let us note first that this action is well defined: if $[x,t] = [x',t']$, then $x' = x + k$ and $t' = t + \tau(k)(x)$, thus $t' + s = t + s + \tau(k)(x)$ and $[x', t' + s] = [x, t + s]$.

Next, for the ones familiar with diffeology, note that this construction is not exactly an associated bundle to the covering $\mathrm{pr} : \mathbf{R} \to T$ as described in [PIZ13, §8.16], so we shall prove directly that it is indeed an $(\mathbf{R},+)$ principal bundle, by applying the criterion established in [PIZ13, §8.11]. Let us then check that the following action map is an induction,

$$\begin{array}{rccc} F: & Y \times \mathbf{R} & \to & Y \times Y \\ & (y,t) & \mapsto & (y, t_Y(y)) \end{array}$$

**(a)** First of all F is injective: $F(y,t) = F(y',t')$ means $y = y'$ and $t_Y(y) = t'_Y(y)$. Let $y = [x,s]$, then $t_Y(y) = t'_Y(y)$ implies $[x, t+s] = [x, t'+s]$, i.e., there exist $k \in K$ such that $x = x + k$ and $t' + s = t + s + \tau(k,t)$. But $x = x + k$ implies $k = 0$, and then, since $\tau(0,x) = 0$, $t + s = t' + s$ implies $t = t'$.

**(b)** Now,

$$\mathrm{F}(\mathrm{Y}\times\mathbf{R}) = \{(y,y')\in \mathrm{Y}\times\mathrm{Y} \mid \pi(y)=\pi(y')\};$$

Consider a plot in $\mathrm{F}(\mathrm{Y}\times\mathbf{R})$, that is, a plot $r\mapsto(y_r,y'_r)$ in $\mathrm{Y}\times\mathrm{Y}$ such that $\pi(y_r)=\pi(y'_r)$, for all $r$. Since the map $\mathbf{R}\times\mathbf{R}\to\mathrm{Y}=\mathbf{R}\times_\tau\mathrm{Y}$ is a subduction, there are, locally everywhere, two smooth parametrizations $r\mapsto(x_r,t_r)$ and $r\mapsto(x'_r,t'_r)$ in $\mathbf{R}\times\mathbf{R}$ such that: $y_r=[x_r,t_r]$ and $y'_r=\left[x'_r,t'_r\right]$, with $\pi(y_r)=\pi(y'_r)=[x_r]$. Thus, there exists a smooth map $r\mapsto k_r$ such that $x'_r=x_r+k_r$, but since $\mathrm{K}\subset\mathbf{R}$ is discrete, locally everywhere: $k_r=k$ and $x'_r=x_r+k$. Hence,

$$\begin{aligned} y'_r = \left[x_r+k,t'_r\right] = \left[x_r,t'_r-\tau(k,x_r)\right] &= \left[x_r,t_r+(t'_r-t_r-\tau(k,x_r))\right] \\ &= [x_r,t_r+s_r] \ \text{ with } \ s_r=t'_r-t_r-\tau(k,x_r). \end{aligned}$$

Then, $y'_r=[x_r,t_r+s_r]=(s_r)_{\mathrm{Y}}(y_r)=\mathrm{F}(y_r,s_r)$, and $r\mapsto s_r$ is a smooth parametrization in $\mathbf{R}$. Therefore, F is an induction and that achieves the proof that $\pi:\mathrm{Y}=\mathbf{R}\times_\tau\mathbf{R}\to\mathrm{T}=\mathbf{R}/\mathrm{K}$ is a $(\mathbf{R},+)$ principal fiber bundle.

**(3)** Now, let us prove that H is a homomorphism. Let $\pi:\mathrm{Y}=\mathbf{R}\times_\tau\mathbf{R}\to\mathrm{T}=\mathbf{R}/\mathrm{K}$ and $\pi':\mathrm{Y}'=\mathbf{R}\times_{\tau'}\mathbf{R}\to\mathrm{T}=\mathbf{R}/\mathrm{K}$. The proof that $\mathrm{H}(\tau+\tau')=\mathrm{H}(\tau)+\mathrm{H}(\tau')$ will be done in two steps. Let us introduce first the following notations:

- $\mathrm{Y}=\mathrm{Y}_\tau=\mathbf{R}\times_\tau\mathbf{R}:=(\mathbf{R}\times\mathbf{R})/\mathrm{K}$ with the action of K defined by $\tau$.
- $\mathrm{Y}'=\mathrm{Y}_{\tau'}=\mathbf{R}\times_{\tau'}\mathbf{R}:=(\mathbf{R}\times\mathbf{R})/\mathrm{K}$ with the action of K defined by $\tau'$.
- $\mathrm{Y}_{\tau,\tau'}=\mathbf{R}\times_{\tau,\tau'}\mathbf{R}^2:=(\mathbf{R}\times\mathbf{R}^2)/\mathrm{K}$ with the action of K on $\mathbf{R}\times\mathbf{R}^2$ associated with the pair $(\tau,\tau')$, defined, for all $k\in\mathrm{K}$ and $(t,t')\in\mathbf{R}^2$, by:

$$k_{\mathbf{R}\times\mathbf{R}^2}(x,t,t') = (x+k,t+\tau(k,x),t'+\tau'(k,x)).$$

- $\mathrm{Y}_{\tau+\tau'}=\mathbf{R}\times_{\tau+\tau'}\mathbf{R}:=(\mathbf{R}\times\mathbf{R})/\mathrm{K}$ with the action of K defined by $\tau+\tau'$.

We write $[x,t]_\tau$, $[x,t']_{\tau'}$ and $[x,t,t']_{\tau,\tau'}$ the elements of $\mathrm{Y}_\tau$, $\mathrm{Y}_{\tau'}$ and $\mathrm{Y}_{\tau,\tau'}$. The first step consists to prove that

$$\mathrm{Y}\otimes\mathrm{Y}'\simeq\mathrm{Y}_{\tau,\tau'}.$$

Indeed, consider the diagrams

$$\begin{array}{ccc} (x,t,t') & \overset{j}{\longmapsto} & ((x,t),(x,t')) \\ {\scriptstyle\pi_{\tau,\tau'}}\big\downarrow & & \big\downarrow{\scriptstyle\pi_\tau\otimes\pi_{\tau'}} \\ [x,t,t']_{\tau,\tau'} & \underset{\underline{j}}{\longmapsto} & ([x,t]_\tau,[x,t']_{\tau'}) \end{array} \qquad \begin{array}{ccc} \mathbf{R}\times\mathbf{R}^2 & \overset{j}{\hookrightarrow} & (\mathbf{R}\times\mathbf{R})\times(\mathbf{R}\times\mathbf{R}) \\ {\scriptstyle\pi_{\tau,\tau'}}\big\downarrow & & \big\downarrow{\scriptstyle\pi_\tau\otimes\pi_{\tau'}} \\ \mathrm{Y}_{\tau,\tau'} & \underset{\underline{j}}{\longrightarrow} & \mathrm{Y}_\tau\otimes\mathrm{Y}_{\tau'} \end{array}$$

The map $j$ is an induction, that is a diffeomorphism onto its image equipped with the subset diffeology. Then, it descends to the quotient in $\underline{j}:[x,t,t']_{\tau,\tau'}\mapsto([x,t]_\tau,[x,t]_{\tau'})$, where it is a smooth bijection onto its image. Since $j$ is an induction, the inverse $\underline{j}^{-1}$ is smooth, and then $\underline{j}$ a strict map [PIZ13, §1.54], i.e. a diffeomorphism onto its image.

The second step consists to show that the quotient $(\mathrm{Y}\otimes\mathrm{Y}')/\overline{\mathbf{R}}$, which is equivalent to $\mathrm{Y}_{\tau,\tau'}/\overline{\mathbf{R}}$, with $\overline{\mathbf{R}}$ acts according to $s_{\tau,\tau'}[x,t,t']_{\tau,\tau'}=[x,t+s,t'-s]_{\tau,\tau'}$ is realized by the projection

$$\mathrm{pr}_{\tau,\tau'}:\mathrm{Y}_{\tau,\tau'}\to\mathrm{Y}_{\tau+\tau'} \ \text{ with } \ \mathrm{pr}_{\tau,\tau'}:\left[x,t,t'\right]_{\tau,\tau'}\mapsto\left[x,t+t'\right]_{\tau+\tau'}.$$

That achieves to prove that $\mathrm{H}(\tau)+\mathrm{H}(\tau')=\mathrm{H}(\tau+\tau')$.

**(4)** We now prove that

$$\ker(\mathrm{H})=\mathbf{B}^1(\mathrm{K},\mathscr{C}^\infty(\mathbf{R})).$$

A cocycle $\tau$ defines a trivial principal bundle if and only if there is a global smooth section $\boldsymbol{\sigma}:\mathrm{T}\to\mathrm{Y}$. Then, $\mathrm{pr}\circ\boldsymbol{\sigma}:\mathbf{R}\to\mathrm{Y}$ has a smooth lift $x\mapsto(x,\sigma(x))$ in $\mathbf{R}\times\mathbf{R}$ such that $\boldsymbol{\sigma}([x])=[x,\sigma(x)]$. This can be regarded as a consequence of the monodromy theorem [PIZ13, §8.25], since the projection from $\mathbf{R}\times\mathbf{R}$ to its quotient Y is a covering, actually the universal covering. Now, for all $k\in\mathrm{K}$, on the one hand $\boldsymbol{\sigma}([x])=\boldsymbol{\sigma}([x+k])$ gives $[x,\sigma(x)]=[x+k,\sigma(x+k)]$, and on the other hand $[x,\sigma(x)]=[x+k,\sigma(x)+\tau(k,x)]$. Thus, $\sigma(x+k)=\sigma(x)+\tau(k,x)$, that is, $\tau(k,x)=\sigma(x+k)-\sigma(x)$, meaning $\tau=\Delta\sigma$, and therefore $\ker(\mathrm{H})=\mathbf{B}^1(\mathrm{K},\mathscr{C}^\infty(\mathbf{R}))$. In consequence, two maps $\tau$ and $\tau'$ satisfying (II) define equivalent $(\mathbf{R},+)$ principal bundles if and only if there exists a map $\sigma\in\mathscr{C}^\infty(\mathbf{R})$ such that: $\tau'(k,x)=\tau(k,x)+\sigma(x+k)-\sigma(x)$. Therefore, $\mathrm{H}:\mathbf{Z}^1(\mathrm{K},\mathscr{C}^\infty(\mathbf{R}))\to\mathbf{Fl}(\mathrm{T},\mathbf{R})$ descends to the quotient $\mathrm{H}:\mathbf{H}^1(\mathrm{K},\mathscr{C}^\infty(\mathbf{R}))\to\mathbf{Fl}(\mathrm{T},\mathbf{R})$.

**(5)** Finally, let us prove that $\mathrm{H}:\mathbf{Z}^1(\mathrm{K},\mathscr{C}^\infty(\mathbf{R}))\to\mathbf{Fl}(\mathrm{T},\mathbf{R})$ is surjective. Let $\pi:\mathrm{Y}\to\mathrm{T}$ be an $(\mathbf{R},+)$ principal bundle. Consider the pullback of $\pi:\mathrm{Y}\to\mathrm{X}$ by the projection $\mathrm{pr}:\mathbf{R}\to\mathrm{T}$, that is, the first projection $\mathrm{pr}_1:\mathrm{pr}^*(\mathrm{Y})\to\mathbf{R}$, with

$$\mathrm{pr}^*(\mathrm{Y})=\{(x,y)\in\mathbf{R}\times\mathrm{Y}\mid\mathrm{pr}(x)=\pi(y)\}.$$

Since it is a principal fiber bundle over a manifold with a contractible fiber, it has a smooth section and then it is trivial [Hus94]. Let $\Phi:\mathbf{R}\times\mathbf{R}\to\mathrm{pr}^*(\mathrm{Y})$ such an isomorphism from the trivial bundle to $\mathrm{pr}^*(\mathrm{Y})$. Since $\mathrm{pr}_1\circ\Phi=\mathrm{pr}_1$, the isomorphism $\Phi$ writes

$$\Phi(x,t)=(x,\varphi(x,t)),$$

where $\varphi:\mathbf{R}\times\mathbf{R}\to\mathrm{Y}$ is smooth. As an isomorphism of principal fiber bundle, it satisfies $\Phi(s_{\mathbf{R}\times\mathbf{R}}(x,t))=s_{\mathrm{pr}^*(\mathrm{Y})}(\Phi(x,t))$, that is, $\Phi(x,s+t)=(x,\varphi(x,s+t))$, and then $s_{\mathrm{pr}^*(\mathrm{Y})}(x,\varphi(x,t))=(x,s_{\mathrm{Y}}(\varphi(x,t)))$. Thus,

$$\varphi(x,s+t)=s_{\mathrm{Y}}(\varphi(x,t)),\ \text{ and then }\ \varphi(x,t)=t_{\mathrm{Y}}(\varphi(x,0)).$$

Let us rename

$$\varphi(x):=\varphi(x,0).$$

The map $\varphi:\mathbf{R}\to\mathrm{Y}$ is a smooth lift of pr, that is, $\pi\circ\varphi=\mathrm{pr}$, which is summarized by the following diagram.

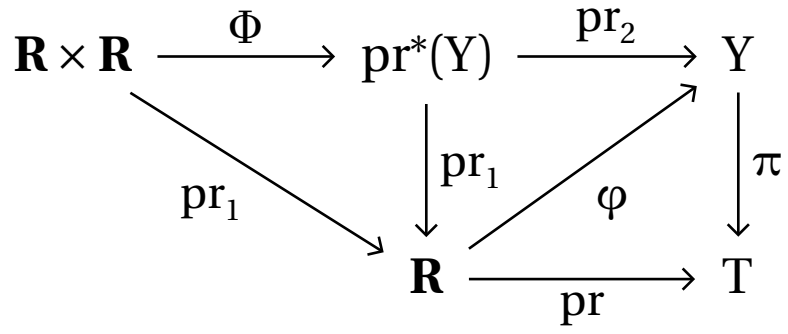

The map $\Phi:(x,t)\mapsto(x,t_{\mathrm{Y}}(\varphi(x)))$ being an isomorphism from $\mathbf{R}\times\mathbf{R}$ to $\mathrm{pr}^*(\mathrm{Y})$, and since $\pi$ and pr are subductions, the second projection $\mathrm{pr}_2:\mathrm{pr}^*(\mathrm{Y})\to\mathrm{Y}$ is a subduction.

Thus, Y is equivalent to the quotient of $\mathbf{R}\times\mathbf{R}$ by the relation $(x,t)\sim(x',t')$ if and only if $\mathrm{pr}_2\circ\Phi(x,t)=\mathrm{pr}_2\circ\Phi(x',t')$, that is, $t_{\mathrm{Y}}(\varphi(x))=t'_{\mathrm{Y}}(\varphi(x'))$. This implies:

(1) There exist $k\in\mathrm{K}$ such that $x'=x+k$.

(2) There exists a map $\tau : \mathrm{K} \times \mathbf{R} \to \mathbf{R}$ such that

$$\varphi(x+k) = \tau(k,x)_{\mathrm{Y}}(\varphi(x)).$$

The map $\tau$ satifies the *cocycle condition*:

$$\tau(k+k',x) = \tau(k,x+k') + \tau(k',x).$$

Therefore, $(x',t') \sim (x,t)$ if and only if

$$x' = x+k \quad \text{and} \quad t' = t + \tau(k,x).$$

This cocycle $\tau \in \mathscr{C}^\infty(\mathrm{K},\mathbf{R})$ lifts the action on K on $\mathbf{R}$ to $\mathbf{R}\times\mathbf{R}$ according to Eq. (E). And since $\mathrm{pr}_2 \circ \Phi$ is a subduction, Y is isomorphic to the quotient of $\mathbf{R}\times\mathbf{R}$ by this action:

$$\mathrm{Y} \simeq \mathbf{R} \times_\tau \mathbf{R} = (\mathbf{R}\times\mathbf{R})/\mathrm{K}.$$

This achieves the proof that the induced homomorphism $\mathrm{H} : \mathsf{H}^1(\mathrm{K}, \mathscr{C}^\infty(\mathbf{R})) \to \mathbf{Fl}(\mathrm{T},\mathbf{R})$ is surjective. As shown in Part (4), it is also injective, and hence an isomorphism. ►

**II.3. Irrational Tori and Small Divisors: the Group $\mathbf{Fl}(\mathrm{T}_\alpha,\mathbf{R})$.**

We shall now explore the geometry delivered by the group $\mathbf{Fl}(\mathrm{T}_\alpha,\mathbf{R})$. This will illustrate the sensibility of the geometry (i.e., diffeology) of the quotient $\mathrm{T}_\mathrm{K} = \mathbf{R}/\mathrm{K}$, captured by the group $\mathbf{Fl}(\mathrm{T}_\mathrm{K},\mathbf{R})$, for the key example of $\mathrm{K} = \mathbf{Z} + \alpha\mathbf{Z}$, where $\alpha \in \mathbf{R} - \mathbf{Q}$ and $\mathbf{R}/\mathrm{K} = \mathrm{T}_\alpha$, leading to distinct behaviors depending on the arithmetic nature of $\alpha$.

Consider the representation of the irrational torus $\mathrm{T}_\alpha = \mathbf{R}/(\mathbf{Z}+\alpha\mathbf{Z})$ as $(\mathbf{R}/\mathbf{Z})/\alpha\mathbf{Z} = \mathrm{T}/\alpha\mathbf{Z}$, where $\mathrm{T} = \mathbf{R}/\mathbf{Z}$, with $n(x) = x+n$, $x \in \mathbf{R}$ and $n \in \mathbf{Z}$, and then $\alpha\mathbf{Z}$ acts on T by $\alpha m[x] = [x+\alpha m]$, where $[x] \in \mathrm{T}$ denotes the class of $x$ modulo $\mathbf{Z}$. Of course, $\mathrm{T} \simeq \mathrm{S}^1 \subset \mathbf{C}$ by the identification $[x] \mapsto z$, with $z = e^{i2\pi x}$, and we will use indifferently $[x]$ or $z$ depending on context.

Let's start by describing the cocycle $\tau$ in this particular case.

**(A) The Cocycle $\tau$.**

Let us denote by $\mathrm{pr} : \mathrm{T} \to \mathrm{T}_\alpha$ the projection, and let $\pi : \mathrm{Y} \to \mathrm{T}_\alpha$ be an $(\mathbf{R},+)$ principal fiber bundle. The pullback $\mathrm{pr}_1 : \mathrm{pr}^*(\mathrm{Y}) \to \mathrm{T}$ is an $(\mathbf{R},+)$ principal fiber bundle over a manifold $\mathrm{T} \simeq \mathrm{S}^1$, thus, it is trivial: $\mathrm{pr}^*(\mathrm{Y}) \simeq \mathrm{T}\times\mathbf{R}$. And the previous reconstruction of Y, for K acting on $\mathbf{R}\times\mathbf{R}$, can be mimicked in this situation with $\mathbf{Z}$ acting on $\mathrm{T}\times\mathbf{R}$ through a cocycle $\tau : \mathbf{Z} \to \mathscr{C}^\infty(\mathrm{T},\mathbf{R})$ satisfying

$$\tau(m+m')([x]) = \tau(m)\big([x+m'\alpha]\big) + \tau(m')([x]).$$

**Proposition.** *Every cocycle $\tau$ is uniquely defined by a function $f \in \mathscr{C}^\infty(\mathrm{T},\mathbf{R})$, and every function $f \in \mathscr{C}^\infty(\mathrm{T},\mathbf{R})$ uniquely defines a cocycle $\tau$ via the formula, for all positive integers $m$:*

$$\tau(m)([x]) = \sum_{k=0}^{m-1} f([x+k\alpha]), \quad \textit{with} \ \ f = \tau(1),\ \tau(0) = 0,$$

$$\textit{and} \ \ \tau(-m)([x]) = -\sum_{k=0}^{m-1} f([x+(k-m)\alpha]), \qquad (\clubsuit)$$

Thus, the cohomology group $\mathbf{Fl}(\mathrm{T}_\alpha, \mathbf{R})$ is a quotient of $\mathscr{C}^\infty(\mathrm{T}, \mathbf{R})$, which we will now determine.

**(B) The Action of Z on** $\mathrm{S}^1 \times \mathbf{R}$**.**

Thanks to the explicit expression of the cocycle $\tau$ in equation (♣), the action of $\mathbf{Z}$ on $\mathrm{S}^1 \times \mathbf{R}$ writes, with $m > 0$:

$$\underline{m}_{\mathrm{S}^1\times\mathbf{R}}(z,t) = \left( ze^{i2\pi m\alpha}, t + \sum_{k=0}^{m-1} f\left(ze^{i2\pi m\alpha}\right)\right)$$
$$\underline{-m}_{\mathrm{S}^1\times\mathbf{R}}(z,t) = \left( ze^{-i2\pi m\alpha}, t - \sum_{k=0}^{m-1} f\left(ze^{i2\pi(k-m)\alpha}\right)\right),$$

where we use the complex notation $z = e^{i2\pi x}$ for $[x]$. Let us introduce now the following diffeomorphism $h \in \mathrm{Diff}(\mathrm{S}^1 \times \mathbf{R})$

$$h : z \mapsto \left(ze^{i2\pi\alpha}, t + f(z)\right) \text{ and } h^{-1}(z,t) = \left(ze^{-i2\pi\alpha}, t - f\left(ze^{-i2\pi\alpha}\right)\right).$$

For all $m >$, let us denote by $h^m$ the composite $m$-times of $h$, by $h^{-m}$, the composite $m$-times of $h^{-1}$, and $h^0 = \mathbf{1}_{\mathrm{S}^1\times\mathbf{R}}$. Then, the action of $\mathbf{Z}$ on $\mathrm{S}^1 \times \mathbf{R}$ writes, for all $m \in \mathbf{Z}$

$$\underline{m}_{\mathrm{S}^1\times\mathbf{R}}(z,t) = h^m(z,t).$$

Thus, the space $\mathrm{Y} = \mathrm{S}^1 \times_\tau \mathbf{R}$ is the quotient of $\mathrm{S}^1 \times \mathbf{R}$ by the $\mathbf{Z}$-iterations of the diffeomorphism $h$, what we denote by:

$$\mathrm{Y} : (\mathrm{S}^1 \times \mathbf{R})/\langle h\rangle.$$

**(C) The Cohomological Equation.**

A function $f \in \mathscr{C}^\infty(\mathrm{T}, \mathbf{R})$ defines a coboundary if and only if there exists $g \in \mathscr{C}^\infty(\mathrm{T}, \mathbf{R})$ satisfying

$$f([x]) = g([x+\alpha]) - g([x]).$$

Let $\mathrm{F}, \mathrm{G} \in \mathscr{C}^\infty_{\mathrm{per}}(\mathbf{R})$ denote the 1-periodic functions representing $f$ and $g$, respectively. The above equation is then equivalent to

$$\mathrm{F}(x) = \mathrm{G}(x+\alpha) - \mathrm{G}(x).$$

This equation in G, the *cohomological equation*, is central to the study of dynamical systems, particularly circle rotations, with deep historical roots in problems involving small divisors (see, e.g., Arnold [Arn65, Arn80], Moser [Mos66], Herman [Her79], Katok and Hasselblatt [KH95], and Ghy's survey [Ghy07]).

**(D) The Decomposition of** $\mathbf{Fl}(\mathrm{T}_\alpha, \mathbf{R})$**.**

Let $\mathscr{C}^\infty_{\mathrm{per},0}(\mathbf{R}) \subset \mathscr{C}^\infty_{\mathrm{per}}(\mathbf{R})$ be the vector subspace of 1-periodic smooth real functions with mean value 0 :

$$\mathscr{C}^\infty_{\mathrm{per},0}(\mathbf{R}) = \left\{ \mathrm{F} \in \mathscr{C}^\infty_{\mathrm{per}}(\mathbf{R}) \;\middle|\; \int_0^1 \mathrm{F}(x)\,dx = 0 \right\}.$$

Consider the linear map

$$\Delta_\alpha : \mathscr{C}^\infty_{\mathrm{per}}(\mathbf{R}) \to \mathscr{C}^\infty_{\mathrm{per},0}(\mathbf{R}), \quad \text{with} \quad \Delta_\alpha(\mathrm{G}) = [x \mapsto \mathrm{G}(x+\alpha) - \mathrm{G}(x)],$$

Recalling that $\mathbf{Fl}(\mathrm{T}_\alpha, \mathbf{R})$ is the quotient of $\mathscr{C}^\infty(\mathrm{T}, \mathbf{R})$ (or equivalently $\mathscr{C}^\infty_{\mathrm{per}}(\mathbf{R})$) by the image of the related map $\mathrm{G} \mapsto \Delta_\alpha(\mathrm{G})$, we can analyze this quotient by decomposing $\mathscr{C}^\infty_{\mathrm{per}}(\mathbf{R})$ into its mean value and zero-mean components. Let:

$$\mathrm{F} \mapsto (c, \delta) \quad \text{with} \quad \begin{cases} c = \displaystyle\int_0^1 \mathrm{F}(x)\,dx & \in \mathbf{R}, \\ \delta = \mathrm{F} - \displaystyle\int_0^1 \mathrm{F}(x)\,dx & \in \mathscr{C}^\infty_{\mathrm{per},0}(\mathbf{R}). \end{cases}$$

The mean value $c$ is clearly not affected by the coboundary condition $\Delta_\alpha \mathrm{G}$. The obstructions for the zero-mean part $\delta$ lie precisely in the cokernel of $\Delta_\alpha$ restricted to functions mapping to $\mathscr{C}^\infty_{\mathrm{per},0}(\mathbf{R})$. This leads directly to the isomorphism:

$$\mathbf{Fl}(\mathrm{T}_\alpha, \mathbf{R}) \simeq \mathbf{R} \times \mathrm{coker}(\Delta_\alpha).$$

This decomposition separates each bundle class into two components: the mean value $c$, which we identify as the *drift parameter* (determining the average vertical shift in the bundle construction and related to the flow speed $1/c$), and a zero-mean part $\delta = f - c$, the *fluctuation component*, whose class lies in $\mathrm{coker}(\Delta_\alpha)$. The naturalness of this decomposition is particularly obvious in the expression of the generating diffeomorphism of the $\mathbf{Z}$ action on $\mathrm{S}^1 \times \mathbf{R}$, namely:

$$h(z, t) = \big(ze^{i2\pi\alpha}, t + c + \delta(z)\big),$$

Thanks to this decomposition into a significant product, we now analyze the structure of the bundle $\pi : \mathrm{Y} = (\mathrm{S}^1 \times \mathbf{R})/\langle h \rangle \to \mathrm{T}_\alpha$ in $\mathbf{Fl}(\mathrm{T}_\alpha, \mathbf{R})$, based on the value of the drift parameter $c$.

**(D.1) Non-zero drift: $\mathbf{c \neq 0}$.**

In this case, the total space Y is a manifold, always diffeomorphic to $\mathrm{T}^2$, only the nature of the flow changes depending on the fluctuation $\delta$. The drift $c$ governs the manifold nature of the total space.

**(D.1.a) Case $c \neq 0$ and $\delta = 0$.**

This case describes the subspace $\mathbf{R} \times \{0\} \subset \mathbf{Fl}(\mathrm{T}_\alpha, \mathbf{R})$. The cocycle $\tau$ is not trivial and equivalent to the morphism

$$\tau(m) : z \mapsto mc \quad \text{and} \quad f(z) = \tau(1)(z) = c = \text{const}.$$

The corresponding diffeomorphism generating the quotient is

$$h : (z, t) \mapsto \big(ze^{i2\pi\alpha}, t + c\big).$$

The structure of Y and the $(\mathbf{R}, +)$ dynamics are given by:

**Proposition.** *The total space* $\mathrm{Y} = (\mathrm{S}^1 \times \mathbf{R})/\langle h \rangle$ *is diffeomorphic to the torus* $\mathrm{T}^2$. *The principal bundle* $\pi : \mathrm{Y} = \mathrm{S}^1 \times_\tau \mathbf{R} \to \mathrm{T}_\alpha$ *is isomorphic to the standard fibration* $\pi_\alpha : \mathrm{T}^2 \to \mathrm{T}_\alpha$. *The* $(\mathbf{R}, +)$ *action corresponds, under this isomorphism, to the standard linear flow on* $\mathrm{T}^2$ *run at speed* $1/c$.

This action of $(\mathbf{R}, +)$ on $\mathrm{T}^2$ is also known as the linear flow running at speed $1/c$. It is denoted in the dynamic systems literature by $\mathrm{L}_{1/c}$.

This case, where the class $\delta$ in $\operatorname{coker}(\Delta_\alpha)$ is zero, is particularly significant because it encompasses *all* non-trivial bundles when $\alpha$ is Diophantine. Indeed, in this case, the cohomological equation above is known to be solvable for any 1-periodic functions with zero mean. This solvability implies that $\operatorname{coker}(\Delta_\alpha) = 0$ (cf. [KH95, Thm. 12.2.2]). Consequently, for Diophantine $\alpha$, every class in $\mathbf{Fl}(\mathrm{T}_\alpha, \mathbf{R})$ is represented by a constant function. In other words:

**Theorem.** *For $\alpha$ Diophantine, the group $\mathbf{Fl}(\mathrm{T}_\alpha, \mathbf{R})$ is isomorphic to $\mathbf{R}$. The total space of any non-trivial $(\mathbf{R},+)$-principal bundle over $\mathrm{T}_\alpha$ (corresponding to $c \neq 0$) is diffeomorphic to $\mathrm{T}^2$, and the fibration $\pi$ is equivalent to the standard linear flow on $\mathrm{T}^2$ run at speed $1/c$.*

**(D.1.b) Case $c \neq 0$ and $\delta \neq 0$.**

Since $\delta \neq 0$ this case concerns uniquely numbers $\alpha$ that are not Diophantine.[6] They are irrational numbers that are very well approximated by rationals. Precisely, $\alpha$ is *non-Diophantine* if it is irrational and for every integer $k \geq 1$, there exists infinitely many pairs of integers $(m, n)$ with $n \geq 1$, such that:[7]

$$0 < |n\alpha - m| < \frac{1}{n^k}.$$

For non-Diophantine $\alpha$ the structure of $\mathbf{Fl}(\mathrm{T}_\alpha, \mathbf{R})$ is made explicit by this theorem:

**Theorem.** *For non-Diophantine numbers $\alpha$, unlike the Diophantine case where $\operatorname{coker}(\Delta_\alpha)$ reduces to $\{0\}$, the cokernel is infinite-dimensional:*

$$\dim(\operatorname{coker}(\Delta_\alpha)) = \infty.$$

*Consequently*

$$\dim(\mathbf{Fl}(\mathrm{T}_\alpha, \mathbf{R})) = 1 + \infty.$$

The infinite dimensionality of $\operatorname{coker}(\Delta_\alpha)$ for non-Diophantine $\alpha$ is a well-known consequence in the specialized field studying small divisor problems. On this general question, see for example the discussions in [Her79, Arn83, KH95, Yoc95, For02]. But for the sake of self-consistency we shall give a proof below.

The geometry of these cases is given by the following theorem:

**Theorem.** *For a cocycle $\tau$ corresponding to a pair $(c, \delta)$ with $c \neq 0$ and $\delta \neq 0$, the space $\mathrm{Y} = \mathrm{S}^1 \times_\tau \mathbf{R}$ is still diffeomorphic to the 2-torus $\mathrm{T}^2$. The transmutation of the action $s[z, t] = [z, t+s]$ on $\mathrm{Y}$ to $\mathrm{T}^2$ is a $(\mathbf{R},+)$ principal fibration which is not conjugate to the standard linear flow of slope $\alpha$, but which shares the same (diffeological) space of orbits (the base space of the fiber bundle) $\mathrm{T}_\alpha$.*

---

[6] There is a convention sometimes found in dynamical systems literature, referring to any non-Diophantine irrational number (equivalently, any irrational with infinite Liouville-Roth irrationality measure) as a Liouville number. This should be distinguished from the stricter definition, often used in number theory requiring $|\alpha - p/q| < 1/q^n$ for all $n \geq 1$, identifies a subset of these numbers. For our purposes, the relevant class is precisely the set of non-Diophantine numbers.

[7] Note that stating existence for infinitely many pairs $(m, n)$ is equivalent to stating existence of at least one pair $(m, n)$ with $n \geq 2$ for each $k$, because if one exists for $k$, one must exist for $k+1$ etc., leading to infinitely many distinct pairs overall as $k$ increases.

We shall see in details in the proof, but the fact that Y is a manifold is a consequence that, when $c \neq 0$, the diffeomorphism $h$ generating the quotient acts freely and properly discontinuously. Then, because we know that $\mathbf{R}^2$ is its universal covering and $\mathbf{Z}^2$ its fundamental group, Y can only be diffeomorphic to $\mathrm{T}^2$.

**(D.2) Zero drift: c = 0.**

In this case, the total space Y is no longer a manifold but the product $\mathrm{T}_\alpha \times \mathbf{R}$ as a set. It is equipped with the usual product diffeology in the trivial case when the fluctuation vanishes, and with an “exotic” diffeology in case of a non-zero fluctuation.

**(D.2.a) Case $c = 0$ and $\delta = 0$.**

This case corresponds to the unique class of the trivial bundle, whose cocycle $\tau$ is cohomologous to the zero-cocycle $\tau(n, z) = 0$. The space $\mathrm{Y} = \mathrm{S}^1 \times_{\tau=0} \mathbf{R} = (\mathrm{S}^1/\mathbf{Z}) \times \mathbf{R} = \mathrm{T}_\alpha \times \mathbf{R}$ is not a manifold. The projection $\pi$ is the projection on the first factor and the action of $\mathbf{R}$ is the translation by any number on the second factor.

**(D.2.b) Case $c = 0$ and $\delta \neq 0$.**

This case requires $\alpha$ to be non-Diophantine, as $\delta$ represents a non-zero class in $\operatorname{coker}(\Delta_\alpha)$. The defining function $f = \delta$ has zero mean ($\int \delta = 0$). This condition guarantees the existence of solutions $g : \mathrm{S}^1 \to \mathbf{R}$ to the cohomological equation $\delta = g \circ \mathrm{R}_\alpha - g$. Although a continuous solution is known to exist under this condition (cf. [KH95]), the crucial point for our diffeological analysis is that since $[\delta] \neq 0$ in $\operatorname{coker}(\Delta_\alpha)$, no such solution $g$ can be smooth.

The existence of a solution $g$ allows us to construct a “principal projection isomorphism”, namely, $\psi : \mathrm{Y} \to \mathrm{T}_\alpha \times \mathbf{R}$ given by $\psi([z, t]) = ([z], t - g(z))$, intertwining the action of $(\mathbf{R}, +)$ and projecting on $\mathbf{1}_{\mathrm{T}_\alpha}$. We can use this bijection to push the diffeology $\mathscr{D}_\mathrm{Y}$ forward onto the product set $\mathrm{T}_\alpha \times \mathbf{R}$, resulting in a diffeological space $(\mathrm{T}_\alpha \times \mathbf{R}, \mathscr{D}'_{\mathrm{prod}})$ which is diffeologically isomorphic to $(\mathrm{Y}, \mathscr{D}_\mathrm{Y})$ and carries the projection $\mathrm{pr}_1$ as its bundle structure.

This bundle $(\mathrm{T}_\alpha \times \mathbf{R}, \mathscr{D}'_{\mathrm{prod}}) \to \mathrm{T}_\alpha$ represents the class $[\delta] \in \mathbf{Fl}(\mathrm{T}_\alpha, \mathbf{R})$. Its non-triviality (since $[\delta] \neq 0$) manifests directly in the fact that the canonical zero section $s_0 : \mathrm{T}_\alpha \to \mathrm{T}_\alpha \times \mathbf{R}$, $s_0([z]) = ([z], 0)$, is *not smooth* with respect to the pushforward diffeology $\mathscr{D}'_{\mathrm{prod}}$. As shown in the proof, the smoothness of $s_0$ is equivalent to the smoothness of $g$, which fails in this case.

Thus, although set-theoretically equivalent to the trivial bundle in the category of principal projections, the bundle corresponding to $(c = 0, \delta \neq 0)$ is smoothly non-trivial. Its diffeological structure, whether viewed on Y or pushed forward to $\mathrm{T}_\alpha \times \mathbf{R}$, reflects the non-smooth nature of the solution $g$ to the cohomological equation.

**Proof.** We will only prove what cannot be immediately deduced from the text. In particular, parts (A), (B) and (C) require no more proof than is explicitly written.

For the part (D), we shall inspect every statement closely.

(D.1.a) Case $c \neq 0, \delta = 0$. Here $f$ is cohomologous to the constant $c$. The action of $\mathbf{Z}$ on $S^1 \times \mathbf{R}$ is given by $m : (z, t) \mapsto (ze^{i2\pi m\alpha}, t + mc)$. The following projection

$$\mathrm{pr} : S^1 \times \mathbf{R} \to T^2 \text{ with } \mathrm{pr} : \begin{pmatrix} z \\ t \end{pmatrix} \mapsto \begin{pmatrix} z_1 = e^{i2\pi t/c} \\ z_2 = \bar{z}\, e^{i2\pi t\alpha/c} \end{pmatrix}$$

realizes the quotient $(S^1 \times \mathbf{R})/\mathbf{Z}$. The action $s : (z, t) \mapsto (z, t + s)$ is transmuted to

$$s_{T^2} \begin{pmatrix} z_1 \\ z_2 \end{pmatrix} = \begin{pmatrix} e^{i2\pi(t+s)/c} \\ \bar{z}\, e^{i2\pi(t+s)\alpha/c} \end{pmatrix} = \begin{pmatrix} z_1 e^{i2\pi s/c} \\ z_2 e^{i2\pi s\alpha/c} \end{pmatrix}.$$

This is the linear flow on $T^2$ of slope $\alpha$ run at speed $1/c$.

For the Theorem in (D.1.a): If $\alpha$ is Diophantine, the equation $\phi = \Delta_\alpha g$ has a smooth solution $g$ for any smooth $\phi$ with zero mean [KH95, Thm. 12.2.2]. Thus $\mathrm{coker}(\Delta_\alpha) = \{0\}$. The decomposition $\mathbf{Fl}(T_\alpha, \mathbf{R}) \simeq R \times \{0\} \cong R$ follows. All non-trivial bundles ($c \neq 0$) fall into this case, have total space $T^2$, and correspond to scaled linear flows.

(D.1.b) Case $c \neq 0, \delta \neq 0$. This requires $\alpha$ to be non-Diophantine, since $\delta \neq 0$ in the cokernel. The generator is $h(z, t) = (ze^{i2\pi\alpha}, t + f(z))$, where $f = c + \delta$.

The action is free as $\alpha$ is irrational. Let us prove that the action is properly discontinuous. Let

$$S_n(z) = \sum_{k=0}^{n-1} f(ze^{i2\pi k\alpha}) = nc + D_n(z), \text{ where } D_n(z) = \sum_{k=0}^{n-1} \delta(ze^{i2\pi k\alpha}).$$

Since $\delta$ is smooth, it is bounded. Although the cohomological equation $\delta = g \circ R_\alpha - g$ may not admit a continuous solution $g$ for non-Diophantine $\alpha$ (and thus $D_n(z)$ might not be bounded), the sums grow sub-linearly. Since $\delta$ has mean zero, the sequence of Birkhoff averages converges uniformly to zero:

$$\lim_{n\to\infty} \frac{D_n(z)}{n} = \int_{S^1} \delta(z)\, d\mu(z) = 0.$$

Consequently, $D_n(z) = o(n)$. The height of $h^n(z, t)$ is $t + S_n(z) = t + nc + D_n(z) = t + n(c + \varepsilon_n(z))$, where $\varepsilon_n(z) \to 0$ as $n \to \infty$. Since $c \neq 0$, $|t + S_n(z)| \to \infty$ as $|n| \to \infty$, uniformly for $(z, t)$ in any compact set K. Since $c \neq 0$ and $D_n(z)$ is uniformly bounded, $|t + S_n(z)| \to \infty$ as $|n| \to \infty$, uniformly for $(z, t)$ in any compact set K. For any compact K, $h^n(K) \cap K = \varnothing$ for large enough $|n|$. The action is properly discontinuous. Therefore, the quotient $Y = (S^1 \times \mathbf{R})/\langle h \rangle$ is a smooth Hausdorff manifold. We have seen above that its universal cover is $\mathbf{R}^2$ and its fundamental group is $\mathbf{Z}^2$, whatever the quotient is a manifold or not, since manifolds constitute a full and faithful subcategory of diffeological spaces. Therefore, as in case (D.1.a), Y is diffeomorphic to $T^2$. This proves the first part of the Theorem/Proposition in (D.1.b). The second part, that the resulting flow $s \cdot [z, t] = [z, t + s]$ on $Y \cong T^2$ is not smoothly conjugate to the standard linear flow when $\delta \neq 0$, is a deeper result from dynamical systems which we accept here (see e.g., Herman [Her79]).

The result concerning the dimension, namely $\dim(\mathbf{Fl}(T_\alpha, \mathbf{R})) = 1 + \infty$, is proved separately below (see Proof II).

(D.2.a) Case $c = 0, \delta = 0$. The isomorphism $H : \mathbf{H}^1(\mathbf{Z}, \mathscr{C}^\infty(S^1, \mathbf{R})) \to \mathbf{Fl}(T_\alpha, \mathbf{R})$ sends $(0, 0)$ to the trivial bundle $\mathrm{pr}_1 : T_\alpha \times \mathbf{R} \to T_\alpha$.

(D.2.b) Case $c=0, \delta \neq 0$. Again, this requires $\alpha$ to be non-Diophantine. The generating function is $f=\delta$, with $\int \delta = 0$ but $[\delta] \neq 0$ in $\mathrm{coker}(\Delta_\alpha)$.

Since $\int \delta = 0$, the equation $\delta(x) = g(x+\alpha) - g(x)$ admits a unique (up to a constant) solution $g : \mathrm{T} \to \mathbf{R}$. However, because $[\delta] \neq 0$ in the smooth cokernel $\mathrm{coker}(\Delta_\alpha)$, this solution $g$ cannot be smooth.

Using this function $g$, define the map $\psi : \mathrm{Y} \to \mathrm{T}_\alpha \times \mathbf{R}$ by $\psi([z,t]) = ([z], t - g(z))$. This map is a bijection, and it intertwines the $\mathbf{R}$-actions and projections, making it a principal projection isomorphism.

Let $\mathscr{D}_\mathrm{Y}$ be the quotient diffeology on $\mathrm{Y} = (\mathrm{S}^1 \times \mathbf{R})/\langle h \rangle$. Let $\mathscr{D}'_{\mathrm{prod}}$ be the pushforward diffeology on the set $\mathrm{T}_\alpha \times \mathbf{R}$ induced by $\psi$. By construction, $\psi : (\mathrm{Y}, \mathscr{D}_\mathrm{Y}) \to (\mathrm{T}_\alpha \times \mathbf{R}, \mathscr{D}'_{\mathrm{prod}})$ is a diffeological isomorphism.

The bundle $(\mathrm{T}_\alpha \times \mathbf{R}, \mathscr{D}'_{\mathrm{prod}}) \to \mathrm{T}_\alpha$ represents the class $[\delta]$. To check if this bundle is smoothly trivial, we examine its canonical zero section $s_0 : \mathrm{T}_\alpha \to \mathrm{T}_\alpha \times \mathbf{R}$, defined by $s_0([z]) = ([z], 0)$.

The section $s_0$ is smooth (as a map into the pushforward diffeology $\mathscr{D}'_{\mathrm{prod}}$) if and only if the composition $\phi \circ s_0 : \mathrm{T}_\alpha \to \mathrm{Y}$ is smooth (a plot in $\mathscr{D}_\mathrm{Y}$), where $\phi = \psi^{-1}$. We have $\phi([z], t) = \big[z, g(z) + t\big]$, so $(\phi \circ s_0)([z]) = \phi([z], 0) = \big[z, g(z)\big]$. Let $s = \phi \circ s_0$ be this section.

The map $s : \mathrm{T}_\alpha \to \mathrm{Y}$ is smooth if and only if its lift $\tilde{s} : \mathrm{S}^1 \to \mathrm{S}^1 \times \mathbf{R}$, given by $\tilde{s}(z) = (z, g(z))$, is smooth when composed with plots into $\mathrm{S}^1$. This requires the map $z \mapsto g(z)$ to be smooth.

Since $g$ is not smooth in this case, the section $s = \phi \circ s_0$ is not smooth, and consequently the zero section $s_0$ is not smooth with respect to the pushforward diffeology $\mathscr{D}'_{\mathrm{prod}}$.

Therefore, the bundle represented by $(c=0, [\delta] \neq 0)$ is smoothly non-trivial, confirming $[\delta] \neq 0$ in $\mathbf{Fl}(\mathrm{T}_\alpha, \mathbf{R})$. The non-triviality is captured by the non-standard smooth structure $\mathscr{D}'_{\mathrm{prod}}$ placed on the product set $\mathrm{T}_\alpha \times \mathbf{R}$.

The proof of the infinite-dimensionality of the cokernel for non-Diophantine $\alpha$ is a standard result in the theory of small divisors. For the sake of completeness, a detailed proof using Fourier analysis is provided in the Appendix. ►

## III. Conclusion

The irrational torus, an object born from the need to build geometric models for the quasiperiodic properties of quasicrystals, has proven to be a remarkably fertile ground for pure mathematical exploration. Its study has consistently revealed that the diffeological framework is capable of detecting deep algebraic and arithmetic information that is invisible to classical topology. This paper represents the latest step in this long-term program.

The progression of insights has revealed an increasingly fine hierarchy of geometric structure. The initial study uncovered its non-trivial homotopy groups, a purely algebraic feature. This was followed by the discovery that the group of components of its diffeomorphism group, $\pi_0(\mathrm{Diff}(\mathrm{T}_\alpha))$, is a subtle algebraic invariant that distinguishes whether the defining slope $\alpha$ is a quadratic irrationality [DIZ83, IZL90].

The group of flows, $\mathbf{Fl}(T_\alpha, \mathbf{R})$, introduced in this paper, represents a new and deeper layer of this structure. Its decomposition, $\mathbf{Fl}(T_\alpha, \mathbf{R}) \simeq \mathbf{R} \times \operatorname{coker}(\Delta_\alpha)$, when combined with the analysis of the de Rham obstruction, reveals a fundamental stratification of the geometry of flows over $T_\alpha$:

(1) For *Diophantine* $\alpha$, the cokernel vanishes. Thus, $\mathbf{Fl}(T_\alpha, \mathbf{R}) = \mathbf{Fl}^{\bullet}(T_\alpha, \mathbf{R}) \simeq \mathbf{R}$. In this case, every flow over the irrational torus admits a connection 1-form. The geometry is "tame" in the sense that it can be fully described by the infinitesimal theory of connections.

(2) For *non-Diophantine* $\alpha$, the infinite-dimensional $\operatorname{coker}(\Delta_\alpha)$ component represents a vast family of flows whose non-triviality is of a purely analytical nature. These are precisely the bundles that *cannot* be equipped with a connection 1-form. Their structure is so intrinsically tied to the non-solvability of the cohomological equation that they resist any infinitesimal description.

This reveals a new layer of geometric complexity. While it can be shown that any flow over $T_\alpha$ admits a more general diffeological connection (defined as a lifting of paths), the non-Diophantine case provides an infinite-dimensional family of examples where this connection cannot be described infinitesimally by a 1-form. The invariant $\mathbf{Fl}(X, \mathbf{R})$ is thus powerful enough to distinguish these subtle grades of geometric structure. In more extreme cases, such as for non-trivial bundles over the smoothly contractible irrational cone $D_\alpha$, the obstruction is even stronger, precluding the existence of *any* smooth connection whatsoever [**?**].

This consistent pattern—from homotopy, to diffeomorphism groups, and now to the group of flows—demonstrates the profound benefit of a geometric approach to these singular objects. The strangeness of the physical systems they model is mirrored by the strange, yet mathematically rich, properties of their geometric counterparts. Diffeology provides the language and tools to show that what might seem like a "pathological" quotient space is, in fact, a sophisticated geometric object that retains a precise memory of the algebraic and arithmetic data used to construct it.

## Further Remarks and Future Directions

**1. Historical Context.** This paper represents an expanded and detailed account of results originally outlined in the 1986 Comptes Rendus note [PIZ86]. The renewed interest in the interplay between Noncommutative Geometry and Diffeology motivates this wider publication of those results, now presented with full proofs and a more developed conceptual framework.

**2. The Nature of the Obstruction.** Unlike classical $S^1$-bundle theory for manifolds, classified by $H^2(M, \mathbf{Z})$, the group $\mathbf{Fl}(X, \mathbf{R})$ captures finer *analytical obstructions*. For $T_\alpha$, the $\operatorname{coker}(\Delta_\alpha)$ component classifies flows whose non-triviality is invisible to the classical theory of connections and curvature, demonstrating how this diffeological invariant reflects subtle analytical properties.

**3. Rigidity and Symmetry.** The case of $T_{\sqrt{p}}$ (where $p$ is not a perfect square) is particularly interesting. It exhibits both a rich symmetry group ($\pi_0(\mathrm{Diff}(T_{\sqrt{p}})) \simeq \{\pm 1\} \times \mathbf{Z}$) and a rigid geometric structure ($\mathbf{Fl}(\mathrm{T}_{\sqrt{p}}, \mathbf{R}) \simeq \mathbf{R}$, since $\sqrt{p}$ is Diophantine). This phenomenon of rigidity constrained by extensive symmetries is a theme that deserves further attention in diffeology.

**4. A Counter-Intuitive Hierarchy.** It is noteworthy that the geometric complexity captured by $\mathbf{Fl}(\mathrm{T}_\alpha, \mathbf{R})$ behaves counter-intuitively with respect to the arithmetic of $\alpha$. It is the non-Diophantine case, where $\alpha$ is exceptionally well-approximated by rationals, that yields the richest bundle structure ($\dim(\mathbf{Fl}) = \infty$). Conversely, the Diophantine case, where $\alpha$ resists rational approximation, leads to an analytical regularity that restricts the geometry significantly ($\mathbf{Fl} \simeq \mathbf{R}$).

**5. Non-Linear Dynamics.** Our study focused on the linear rotation $\mathrm{R}_\alpha$. A natural extension is to consider "exotic irrational tori" of the form $\mathrm{T}_h = \mathrm{S}^1/\langle h \rangle$, where $h$ is a diffeomorphism that is topologically, but not smoothly, conjugate to $\mathrm{R}_\alpha$. The invariant $\mathbf{Fl}(\mathrm{T}_h, \mathbf{R})$ would then be sensitive not just to the rotation number, but to the finer *smooth conjugacy class* of the generating diffeomorphism, further illustrating the connection between diffeological invariants and subtle properties of dynamical systems.

**6. Analogies in Hyperbolic Geometry.** A compelling direction for future research arises from extending the analogy between linear flows on $\mathrm{T}^2$ and geodesic flows on hyperbolic surfaces. The leaf space of the foliation by geodesics asymptotic to a point on the boundary, $\mathrm{X}_u = \Sigma/\mathscr{F}_u$, serves as a hyperbolic analog of the irrational torus. Investigating the structure of $\mathbf{Fl}(\mathrm{X}_u, \mathbf{R})$ could reveal how diffeological invariants capture information about the geometry of the surface $\Sigma$.

**7. Connection to Noncommutative Geometry and Quasifolds.** The non-triviality of $\mathbf{Fl}(\mathrm{X}, \mathbf{R})$ appears deeply connected to the structure of the gauge groupoid $\mathbf{G}_\mathrm{X}$ associated with a quasifold X, as developed in [IZP21]. Exploring the link between this bundle classification and the properties of the groupoid's C*-algebra presents a promising avenue for bridging the results herein with those of noncommutative geometry.

## Appendix: The Infinite-Dimensionality of the Cokernel

We shall prove using Fourier transforms, that, for $\alpha$ non-Diophantine, the cokernel of $\Delta_\alpha : \mathscr{C}^\infty(\mathrm{T}, \mathbf{R}) \to \mathscr{C}^\infty(\mathrm{T}, \mathbf{R})$, with $\Delta_\alpha(g) = g \circ \mathrm{R}_\alpha - g$, where $\mathrm{R}_\alpha$ is the rotation of angle $\alpha$ (that is, the multiplication by $e^{i2\pi\alpha}$), is infinite-dimensional when $\alpha \in \mathbf{R} - \mathbf{Q}$ is not Diophantine. This will justify the notation $\dim(\mathbf{Fl}(\mathrm{T}_\alpha, \mathbf{R})) = 1 + \infty$. The strategy is to construct an infinite sequence of smooth functions $\{f_m\}$ and demonstrate that they cannot lie in the image of $\Delta_\alpha$; specifically, we show that if we assumed $f_m = \Delta_\alpha g_m$ for some smooth $g_m$, the Fourier coefficients of this hypothetical $g_m$ would necessarily violate the rapid decay condition required for smoothness (i.e., membership in the Schwartz space $\mathscr{S}_\mathbf{R}(\mathbf{Z})$), leading to a contradiction.

**(a) Framework.**

Let $\mathscr{C}^\infty(\mathrm{T}, \mathbf{R})$ be the space of smooth, 1-periodic, real-valued functions on $\mathbf{R}$. We identify $\mathrm{T} = \mathbf{R}/\mathbf{Z}$. Any function $g \in \mathscr{C}^\infty(\mathrm{T}, \mathbf{R})$ has a Fourier series expansion

$$g(x) = \sum_{k \in \mathbf{Z}} \hat{g}(k) e^{i2\pi k x}$$

where the Fourier coefficients are given by

$$\hat{g}(k) = \int_0^1 g(x) e^{-i2\pi kx}\, dx.$$

The condition that $g(x)$ is real-valued for all $x$ is equivalent to the symmetry condition on its Fourier coefficients:

$$\hat{g}(k)^* = \hat{g}(-k) \quad \forall k \in \mathbf{Z},$$

where the asterisk denotes the complex conjugate.

The smoothness of $g$ is equivalent to the rapid decay of its Fourier coefficients, meaning the sequence $\hat{g} = (\hat{g}(k))_{k\in\mathrm{Z}}$ belongs to the Schwartz space $\mathscr{S}(\mathbf{Z})$. This means for every $\mathrm{N} \geq 0$, there exists $\mathrm{C}_\mathrm{N} > 0$ such that

$$|\hat{g}(k)| \leq \frac{\mathrm{C}_\mathrm{N}}{(1+|k|)^\mathrm{N}}, \ \forall k \in \mathbf{Z}.$$

Let

$$\mathscr{S}_\mathbf{R}(\mathbf{Z}) = \{\hat{g} \in \mathscr{S}(\mathrm{Z}) : \hat{g}(k)^* = \hat{g}(-k)\}.$$

The Fourier transform $\mathscr{F} : g \mapsto \hat{g}$ establishes a (topological) vector space isomorphism (over R) between $\mathscr{C}^\infty(\mathrm{T}, \mathbf{R})$ and $\mathscr{S}_\mathbf{R}(\mathbf{Z})$. The operator $\Delta_\alpha g(x) = g(x+\alpha) - g(x)$ clearly maps $\mathscr{C}^\infty(\mathrm{T}, \mathbf{R})$ to itself, since if $g$ is smooth and real-valued, so is $\Delta_\alpha g$.

**(b) Action of $\Delta_\alpha$ in Fourier Space.**

Let $h = \Delta_\alpha g$. Its Fourier coefficients are

$$\hat{h}(k) = \lambda_k \hat{g}(k) \ \text{ with } \ \lambda_k = e^{i2\pi k\alpha} - 1, \ \text{ and then } \ \widehat{\Delta_\alpha g}(k) = \lambda_k \hat{g}(k).$$

The operator $\Delta_\alpha$ corresponds via Fourier transform to the multiplication operator

$$\mathrm{M}_\lambda : \hat{g} \mapsto (\lambda_k \hat{g}(k))_{k\in\mathbf{Z}}.$$

We verify that $\mathrm{M}_\lambda$ maps $\mathscr{S}_\mathbf{R}(\mathbf{Z})$ to itself. If $\hat{g} \in \mathscr{S}_\mathbf{R}(\mathbf{Z})$, let $\hat{h} = \mathrm{M}_\lambda \hat{g}$. Then $\hat{h} \in \mathscr{S}(\mathrm{Z})$ because $\mathscr{S}(\mathrm{Z})$ is an algebra under pointwise multiplication and $(\lambda_k)$ is a bounded sequence (thus a multiplier). We check the reality condition:

$$\hat{h}(k)^* = \big(\lambda_k \hat{g}(k)\big)^* = \lambda_k^* \hat{g}(k)^*, \ \text{ and } \ \hat{h}(-k) = \lambda_{-k} \hat{g}(-k).$$

Since $\alpha$ is real, $\lambda_k^* = (e^{i2\pi k\alpha} - 1)^* = e^{-i2\pi k\alpha} - 1 = \lambda_{-k}$. Since $\hat{g} \in \mathscr{S}_\mathbf{R}(\mathbf{Z})$, we have $\hat{g}(k)^* = \hat{g}(-k)$. Therefore, $\hat{h}(k)^* = \lambda_{-k}\hat{g}(-k) = \hat{h}(-k)$. So, $\hat{h} \in \mathscr{S}_\mathbf{R}(\mathbf{Z})$.

The image $\mathrm{im}(\Delta_\alpha)$ corresponds via $\mathscr{F}$ to $\mathrm{im}(\mathrm{M}_\lambda|_{\mathscr{S}_\mathbf{R}(\mathbf{Z})}) \subset \mathscr{S}_\mathbf{R}(\mathbf{Z})$. The cokernel is $\mathrm{coker}(\Delta_\alpha) = \mathscr{C}^\infty(\mathrm{T}, \mathbf{R}) / \mathrm{im}(\Delta_\alpha)$, which is isomorphic (as a real vector space) to:

$$\mathrm{coker}(\Delta_\alpha) \simeq \mathscr{S}_\mathbf{R}(\mathbf{Z}) / \mathrm{im}(\mathrm{M}_\lambda|_{\mathscr{S}_\mathbf{R}(\mathbf{Z})}).$$

**(c) The Non-Diophantine Condition.**

An irrational number $\alpha$ is *not* Diophantine if for every $\nu \geq 2$ and every $\mathrm{C} > 0$, there exists $k \neq 0$ such that

$$\|k\alpha\| < \frac{\mathrm{C}}{|k|^{\nu-1}}.$$

where $\|x\|$ denotes the distance to the nearest integer $\|x\| = \min_{p\in\mathrm{Z}} |x - p|$. The magnitude of $\lambda_k$ is related to $\|k\alpha\|$:

$$|\lambda_k| = |e^{i2\pi k\alpha} - 1| = |e^{i\pi k\alpha}(e^{i\pi k\alpha} - e^{-i\pi k\alpha})| = |1 \cdot 2i \sin(\pi k\alpha)| = 2|\sin(\pi k\alpha)|.$$

Since $|\sin(\pi x)| = \sin(\pi\|x\|)$ and $\sin(x) \approx x$ for small $x$, we have $|\lambda_k| \approx 2\pi\|k\alpha\|$ when $\|k\alpha\|$ is small. This implies that $|\lambda_k|$ can decay faster than any polynomial rate for certain sequences of $k$'s tending to infinity. Specifically, we can construct an infinite sequence $K^+ = \{k_p\}_{p\geq 1}$ of distinct positive integers such that $k_p \to \infty$ as $p \to \infty$ and satisfying:

$$|\lambda_{k_p}| < \frac{1}{k_p^{2p}} \quad \text{for all } p \geq 1.$$

◀ **Justification**. Since $\alpha$ is not Diophantine, for any integer $N \geq 1$, the set $S_N = \{k \in \mathbf{Z}-\{0\} : |\lambda_k| < 1/|k|^N\}$ is infinite. We construct $K^+$ inductively. Choose $k_1 \in S_2$ with $k_1 > 0$. Assume $k_1, \ldots, k_{p-1}$ have been chosen. Since $S_{2p}$ is infinite, we can choose $k_p \in S_{2p}$ such that $k_p > k_{p-1}$. If this $k_p$ is negative, we can use $|k_p|$ instead, since $|\lambda_{-k}| = |\lambda_k|$ and $|-k|^N = |k|^N$, so $S_N$ is symmetric. Thus we can always choose $k_p > k_{p-1} > 0$. This yields the desired sequence $K^+ = \{k_p\}_{p\geq 1}$ of distinct positive integers with $k_p \to \infty$ and $|\lambda_{k_p}| < 1/k_p^{2p}$. ▶

Let $K = K^+ \cup (-K^+)$. This set K is infinite, symmetric ($k \in K \iff -k \in K$), and $0 \notin K$. For any $k \in K$, if $|k| = k_p$, we have $|\lambda_k| = |\lambda_{k_p}| < 1/|k|^{2p}$.

**(d) Constructing Elements in the Cokernel.**

We construct an infinite set of functions $\{f_m\}_{m\geq 1}$ in $\mathscr{C}^\infty(T, \mathbf{R})$ whose equivalence classes $[f_m]$ in $\operatorname{coker}(\Delta_\alpha)$ are linearly independent over $\mathbf{R}$.

**(d.1) Partition the sequence.** We routinely partition the infinite set $K^+$ into infinitely many disjoint infinite subsets $K_m^+ = \{k_{p_j}^{(m)}\}_{j\geq 1}$ for $m = 1, 2, \ldots$. Let $K_m = K_m^+ \cup (-K_m^+)$. The sets $K_m$ ($m \geq 1$) are pairwise disjoint, symmetric, infinite, contain no zero, and their union is K.

**(d.2) Define functions $f_m$ via Fourier coefficients.** For each $m \geq 1$, define the sequence $\hat{f}_m = (\hat{f}_m(k))_{k\in\mathbf{Z}}$ by:

$$\hat{f}_m(k) = \begin{cases} |\lambda_k|^{1/2} & \text{if } k \in K_m \\ 0 & \text{if } k \notin K_m \end{cases}$$

**(d.3) Check $\hat{f}_m \in \mathscr{S}_R(Z)$.** The reality condition holds by the same argument as in the previous version (symmetry of $K_m$ and $|\lambda_k| = |\lambda_{-k}|$).

For the rapid decay, consider $k \in K_m$. Then $k \in K$, so $|k| = k_p$ for some $p \geq 1$ (specifically, $p = p_j^{(m)}$ if $k = \pm k_{p_j}^{(m)}$). We have $|\lambda_k| < 1/|k|^{2p}$. Thus,

$$|\hat{f}_m(k)| = |\lambda_k|^{1/2} < \left(\frac{1}{|k|^{2p}}\right)^{1/2} = \frac{1}{|k|^p}.$$

Since $k \in K_m$ and $|k| \to \infty$, the corresponding index $p = p(k)$ also tends to infinity. For any fixed $N \geq 0$, we have $p > N$ for $|k|$ sufficiently large. Thus, for $|k|$ large enough and $k \in K_m$, $|\hat{f}_m(k)| < 1/|k|^N$. Since $\hat{f}_m(k) = 0$ for $k \notin K_m$, the sequence $\hat{f}_m$ decays faster than any fixed polynomial rate, hence $\hat{f}_m \in \mathscr{S}(\mathbf{Z})$.

Since $\hat{f}_m$ satisfies both conditions, $\hat{f}_m \in \mathscr{S}_\mathbf{R}(\mathbf{Z})$. Let $f_m = F^{-1}(\hat{f}_m)$. Then $f_m \in \mathscr{C}^\infty(T, \mathbf{R})$.

**(d.4) Show $f_m \notin \mathrm{im}(\Delta_\alpha)$.** Assume, for contradiction, that $f_m = \Delta_\alpha g_m$ for some $g_m \in \mathscr{C}^\infty(\mathrm{T}, \mathbf{R})$. This requires

$$\hat{g}_m(k) = \frac{\hat{f}_m(k)}{\lambda_k} \text{ (for } k \neq 0\text{) to be in } \mathscr{S}_{\mathrm{R}}(\mathrm{Z}).$$

Consider $k \in \mathrm{K}_m$. Then $k \neq 0$ and $|k| = k_p$ for some $p = p(k)$.

$$|\hat{g}_m(k)| = \left|\frac{\hat{f}_m(k)}{\lambda_k}\right| = \frac{|\lambda_k|^{1/2}}{|\lambda_k|} = \frac{1}{|\lambda_k|^{1/2}}.$$

Since $|\lambda_k| < 1/|k|^{2p}$ for $k \in \mathrm{K}_m$, we get

$$|\hat{g}_m(k)| > \frac{1}{(1/|k|^{2p})^{1/2}} = \frac{1}{1/|k|^p} = |k|^p.$$

As $|k| \to \infty$ for $k \in \mathrm{K}_m$, the corresponding index $p = p(k)$ also tends to infinity. The magnitude $|\hat{g}_m(k)|$ grows super-polynomially ($|k|^p$ grows faster than any fixed power $|k|^{\mathrm{N}}$ as $p \to \infty$). This contradicts the requirement that $\hat{g}_m \in \mathscr{S}(\mathrm{Z})$. Therefore, $f_m \notin \mathrm{im}(\Delta_\alpha)$.

**(d.5) Linear Independence.** Suppose $\sum_{m=1}^{\mathrm{M}} c_m[f_m] = [0]$ in $\mathrm{coker}(\Delta_\alpha)$ for real coefficients $c_m$. Then $\mathrm{F} = \sum_{m=1}^{\mathrm{M}} c_m f_m \in \mathrm{im}(\Delta_\alpha)$. Let $\hat{\mathrm{G}} \in \mathscr{S}_{\mathbf{R}}(\mathbf{Z})$ be such that $\hat{\mathrm{F}}(k) = \lambda_k \hat{\mathrm{G}}(k)$ for all $k$. For $k \neq 0$, $\hat{\mathrm{G}}(k) = \hat{\mathrm{F}}(k)/\lambda_k$. Consider $k \in \mathrm{K}_{m_0}$ for some $1 \leq m_0 \leq \mathrm{M}$. Then $\hat{\mathrm{F}}(k) = c_{m_0}\hat{f}_{m_0}(k)$ (since the supports $\mathrm{K}_m$ are disjoint). Let $|k| = k_p$ for $p = p(k)$.

$$\hat{\mathrm{G}}(k) = \frac{c_{m_0}\hat{f}_{m_0}(k)}{\lambda_k} = \frac{c_{m_0}|\lambda_k|^{1/2}}{\lambda_k}.$$

$$|\hat{\mathrm{G}}(k)| = |c_{m_0}|\frac{|\lambda_k|^{1/2}}{|\lambda_k|} = |c_{m_0}|\frac{1}{|\lambda_k|^{1/2}}.$$

If $c_{m_0} \neq 0$, then

$$|\hat{\mathrm{G}}(k)| > |c_{m_0}||k|^p.$$

As $|k| \to \infty$ for $k \in \mathrm{K}_{m_0}$, $p = p(k) \to \infty$, so $|\hat{\mathrm{G}}(k)|$ grows super-polynomially. This prevents $\hat{\mathrm{G}}$ from being in $\mathscr{S}(\mathbf{Z})$. Therefore, we must have $c_{m_0} = 0$. Since this holds for each $m_0 = 1, \ldots, \mathrm{M}$, the classes $[f_m]$ are linearly independent over $\mathbf{R}$.

**Conclusion**. We have constructed an infinite set of functions $\{f_m\}_{m\geq 1}$ in $\mathscr{C}^\infty(\mathrm{T}, \mathbf{R})$ whose equivalence classes $[f_m]$ in the cokernel $\mathrm{coker}(\Delta_\alpha) = \mathscr{C}^\infty(\mathrm{T}, \mathbf{R})/\mathrm{im}(\Delta_\alpha)$ are linearly independent over R. Therefore, the dimension of the cokernel of $\Delta_\alpha$ acting on $\mathscr{C}^\infty(\mathrm{T}, \mathbf{R})$ is infinite when $\alpha$ is irrational and not Diophantine.

EINSTEIN INSTITUTE OF MATHEMATICS, THE HEBREW UNIVERSITY OF JERUSALEM, CAMPUS GIVAT RAM, 9190401 ISRAEL

*Email address*: piz@math.huji.ac.il